\documentclass[journal]{IEEEtran}
\usepackage{csquotes}
\usepackage{fancyvrb}
\usepackage[nolist]{acronym}
\usepackage{pifont}
\usepackage{graphicx}
\usepackage[export]{adjustbox}
\usepackage[super]{nth}
\usepackage{subfig}
\usepackage{float}
\usepackage{tabularx}
\usepackage{cite}
\usepackage{makecell}
\usepackage{array}
\usepackage{multirow}
\usepackage{booktabs}
\usepackage{dblfloatfix}
\usepackage{amsmath}
\usepackage{threeparttable}
\usepackage{afterpage} 
\usepackage{wrapfig}
\usepackage{tikz}
\usetikzlibrary{arrows.meta,positioning,fit,shapes.geometric}
\usepackage{helvet}

\usepackage[final]{microtype}
\renewcommand{\arraystretch}{0.8}
\usepackage[dvipsnames]{xcolor}
\newcommand{\hfrads}{\ac{HFRADS}}
\usepackage[hyphens,spaces,obeyspaces]{url}
\newcolumntype{C}{>{\centering\arraybackslash}X} 
\ifCLASSINFOpdf
  
\else

\fi

\begin{document}
\begin{acronym}
\acro {DNN}{deep neural network}
\acro {GAN}{generative adversarial network}
\acro {GNN}{graph neural network}
\acro {CNN}{convolutional neural network}
\acro {DLA}{deep learning accelerator}
\acro {PVA}{programmable vision accelerator}
\acro {VIC}{video image compositor}
\acro {SSD}{single shot detector}
\acro {GPC}{graphics processing cluster}
\acro {SM}{streaming multiprocessors}
\acro {DMA}{direct memory access}
\acro {VLIW}{very long instruction word}
\acro {VPU}{vector processing unit}
\acro {TPU}{tensor processing unit}
\acro {VPI}{Vision Programming Interface}
\acro {AxoNN}{energy-aware execution of neural networks}
\acro {HaX-CoNN}{heterogeneity aware execution of concurrent deep neural networks}
\acro {D-HaX-CoNN}{dynamic heterogeneity aware execution of concurrent deep neural networks}
\acro {CP-CNN}{computational parallelization for CNNs}
\acro {PCCS}{processor-centric contention-aware slowdown model}
\acro {PND}{partial network duplication}
\acro {SMT}{satisfiability modulo}
\acro {SAT}{satisfiability}
\acro {LP}{linear programming}
\acro {Jedi}{Jetson-aware embedded deep learning inference}
\acro {RNN}{recurrent neural network}
\acro {GA}{genetic algorithm}
\acro {H2H}{heterogeneous model to heterogeneous system mapping}
\acro {MaGNAS}{mapping-aware graph neural architecture search}
\acro {AI-SoC}{AI-System-on-Chip}
\acro {LLM}{large language models}
\acro {RL}{Reinforcement Learning}
\acro {PSNR}{peak-signal-to-noise ratio}
\acro {SSIM}{structural similarity ratio}
\acro {MSE}{mean-square error}
\acro {MRI}{magnetic resonance imaging}
\acro {CT}{computed tomography}
\acro {NPU}{neural processing unit}
\acro {FPGA}{field programmable gate array}
\acro {ASIC}{application specific integrated circuit}
\acro {DSP}{digital signal processor}
\acro {PET}{positron emission tomography}
\acro {OCT}{optical coherence tomography}
\acro {GELU}{Gaussian error linear unit}
\acro {LRN}{local response normalization}
\acro {VLA}{vision-language-action model}
\acro {VLM}{vision language model}
\acro {FSD}{full self-driving}
\acro {ERF}{error function}
\acro {MLP}{multilayer perceptron}
\acro {YOLO}{you only look once}
\acro {VGG}{visual geometry group}
\acro {ViT}{Vision Transformer}
\acro {DVFS}{dynamic voltage and frequency scaling}
\acro {EMC}{external memory controller}
\acro {AV}{autonomous vehicle}
\acro {OFA}{optical flow accelerator}
\acro {HFRADS}[{H-FraDS}]{{Heterogeneous Frame Dispatch Scheduling}}

\end{acronym}
\captionsetup[figure]{justification=centering,font=small}
\captionsetup[table]{font=small}

\title{Edge Physical AI Deployment of Vision Transformers on Heterogeneous Edge GPU Targeting Autonomous Vehicles}

\author{Ashiyana~Abdul Majeed,~\IEEEmembership{Member,~IEEE,}
        Mahmoud~Meribout,~\IEEEmembership{Senior Member,~IEEE,}
        Neethu~Joseph, 
        Abel~Kidane Haile,
        and~Mohammad~Abdullah Al Faruque,~\IEEEmembership{Senior Member,~IEEE} 
\thanks{Ashiyana Abdul Majeed, Dr. Mahmoud Meribout, Neethu Joseph, and Abel Kidane Haile are with the Department of Computer and Information Engineering, Khalifa University, Abu Dhabi, UAE (email: 100059454@ku.ac.ae, mahmoud.meribout@ku.ac.ae, 100069410@ku.ac.ae, ku100053692@alumni.ku.ac.ae). Dr. Mohammad Abdullah Al Faruque is with the Department of Electrical Engineering and Computer Science at the University of California, Irvine, CA 92697, USA (email: alfaruqu@uci.edu).}}

\maketitle
 
\begin{abstract}
Physical AI systems, such as autonomous vehicles and intelligent machines, require transformer-based perception models that satisfy stringent edge latency and energy constraints. However, heterogeneous edge-GPU deployment remains limited by underutilized hardware engines and accelerator-incompatible operators, causing fragmented execution and lower throughput per watt. This paper presents \hfrads, a hardware-aware frame scheduling methodology for transformer inference on a recent NVIDIA edge GPU. \hfrads~routes frames across the GPU and dual \ac{DLA} cores using fixed dispatch ratios to improve utilization under latency and power constraints. To enable scheduling, incompatible transformer components are adapted for \ac{DLA} execution by reshaping tensors, approximating \ac{ERF} with $\tanh$, and replacing layer normalization with bounded $\tanh$. The adapted model maintains a 92\% F1 score, with only a 2\% reduction from the original. \Ac{OFA} is further used for inference-side optical-flow estimation. To the best of the authors' knowledge, prior work has not addressed these combined issues. Using Swin Transformer for autonomous-driving perception, \hfrads~Balanced Dispatch (1:2) achieves 125.93 FPS, a 2.36$\times$ speedup over standalone adapted-\ac{DLA} execution, 4.0 FPS/W, and $\approx$24 ms \ac{DLA} latency, satisfying 30 FPS real-time operation; the GPU-\ac{DLA}-\ac{OFA} case achieves a 2.02$\times$ \ac{DLA} throughput speedup.
\end{abstract}
 
 
\IEEEpeerreviewmaketitle

\section{Introduction}

Modern edge-AI systems must execute real-time perception and signal-processing workloads under strict latency, power, and thermal constraints. Applications such as autonomous driving, mobile robotics, intelligent surveillance, industrial inspection, and embedded vision increasingly rely on compact heterogeneous platforms that combine GPU resources with fixed-function engines, including \acfp{DLA}, vision accelerators, video engines, and optical-flow engines. Their practical performance, therefore, depends not only on model accuracy but also on whether the deployment pipeline can exploit the available hardware engines concurrently \cite{baobaid2025edgegpuengines}. Autonomous driving is a representative high-pressure case: perception must support detection and localization under varied environmental conditions within tight computational budgets, since latency or accuracy failures can propagate to prediction, planning, and control. This creates a central trade-off among accuracy, latency, and efficiency, particularly as high-performing models become more computationally demanding.

This work is motivated by the transition from conventional edge AI to edge physical AI, where perception models are no longer isolated classifiers but foundational components inside closed-loop physical agents. In physical AI systems, visual backbones support perception, mapping, tracking, planning, and action selection under real-time constraints. Therefore, the deployment question is not only whether a transformer model is accurate, but whether it can operate continuously on an embedded heterogeneous \ac{AI-SoC} within a fixed power envelope. NVIDIA Jetson AGX Orin is a representative platform for this setting because it combines GPU compute, dual \ac{DLA} engines, \ac{OFA}, high-bandwidth shared memory, and a robotics-oriented software stack \cite{xavier-orin-data}. In this context, hardware-aware transformer adaptation and the \hfrads~scheduler provide the system mechanism for efficiently executing foundational visual perception blocks at the edge.

Previously, \acp{CNN} served as the dominant paradigm for visual perception. Architectures such as ResNet~\cite{he2016resnet}, \ac{VGG}~\cite{simonyan2015vgg}, and \ac{YOLO}~\cite{redmon2016yolo} demonstrated strong performance on standard benchmarks. However, the locality of convolution limits the receptive field's growth with depth, making it difficult to model global context and long-range interactions between objects~\cite{NIPS2016_c8067ad1}, properties critical in complex traffic environments. Achieving full autonomy also requires integrating transformer-based architectures alongside convolutional models for more effective multi-sensor data fusion~\cite{chen2025survey}.
 
Transformer-based models are increasingly attractive for edge perception because they model long-range dependencies and global context. The \ac{ViT} \cite{dosovitskiy2021vit} demonstrated that treating images as sequences of fixed-size patches processed through a standard transformer encoder could match or exceed CNN accuracy on large-scale image classification. Despite this, \ac{ViT}'s global self-attention at a fixed resolution produces a single-scale feature map poorly suited to the multi-scale requirements of object detection and dense prediction, and its quadratic complexity with respect to image resolution is prohibitive for high-resolution automotive cameras~\cite{liu2021swin}. The Swin Transformer~\cite{liu2021swin} addressed both limitations through a hierarchical architecture with shifted-window self-attention, yielding linear complexity and a multi-scale feature pyramid well-matched to autonomous-driving and embedded-vision perception tasks.
 
Modern edge AI platforms integrate dedicated \ac{DLA} engines alongside GPU cores to enhance inference efficiency. A growing body of work addresses edge-aware deployment of \ac{ViT} through quantization, pruning, knowledge distillation, scheduling, and hardware-software co-design~\cite{vit_edge_survey2025, mittal2019jetson}. Scheduling techniques such as Map-and-Conquer~\cite{bouzidi2023mapandconquer} demonstrate that transformer workloads can benefit from heterogeneous mapping across GPU and \ac{DLA} resources. However, they generally assume that the model layers are already compatible with the selected hardware engine \cite{review}. This assumption is restrictive for hierarchical vision transformers, where unsupported operators and tensor layouts can trigger GPU fallback before scheduling can provide its intended benefit. This work addresses that gap by combining transformer compatibility adaptation with hardware-engine-aware scheduling, following the broader hardware-aware deployment philosophy of prior convolutional-model implementations~\cite{mri_pipeline}.

The main contributions are as follows:

\begin{itemize}

    \item \textbf{General accelerator-aware transformer deployment methodology.} A systematic methodology is presented for deploying transformer-based edge-AI models on heterogeneous edge-GPU platforms by combining accelerator-compatibility adaptation with hardware-engine-aware scheduling. The methodology is demonstrated on the Swin Transformer using the latest NVIDIA edge GPU platform for autonomous-driving perception. However, the substitutions apply to transformer architectures that share the same structural incompatibilities. Together, these modifications reduce GPU fallback transitions and improve subsequent scheduling across available hardware engines.

    \item \textbf{Maximum hardware-engine exploration with first frame dispatch scheduling.}
    Five scheduling strategies are designed, implemented, and benchmarked on the recent NVIDIA edge GPU, covering free-running concurrent task execution, single-\ac{DLA} frame dispatch, dual-\ac{DLA} interleaved dispatch, symmetric dual-\ac{DLA} operation, and concurrent GPU-\ac{DLA}-\ac{OFA} execution. To the best of the authors' knowledge, this is the first work to explore frame-based dispatch scheduling on powerful edge GPU platforms such as the NVIDIA Jetson, and the first to integrate the \ac{OFA} as an active concurrent inference unit within a heterogeneous scheduling framework for transformer-based perception. The framework was implemented and evaluated for the first time in urban, crowded road scenes in Abu Dhabi, UAE, to support autonomous-vehicle perception. 

    \item \textbf{Latency and power-efficiency analysis against automotive driving requirements.}
    Per-frame \ac{DLA} perception latency and system-level throughput per watt are reported and evaluated against three automotive latency thresholds: real-time 30 FPS ($\leq$33 ms), urban driving ($\leq$50 ms), and highway driving ($\leq$100 ms).

\end{itemize}

\section{Related Work}

Edge physical AI requires embedded agents to perform perception, reasoning, and action under strict latency, power, memory, and thermal constraints. In this setting, the visual backbone serves as a foundational perceptual block for downstream tasks, including detection, tracking, mapping, forecasting, and planning. Vision transformers are attractive for this role because they capture long-range spatial context. However, their deployment on Jetson-class edge \ac{AI-SoC} platforms requires both accuracy preservation and hardware compatibility. This motivates the proposed Swin Transformer adaptation and \hfrads~scheduler, which connect transformer model design with the heterogeneous GPU, dual-\ac{DLA}, and \ac{OFA} resources of NVIDIA Jetson AGX Orin.

\vspace{-1em}
\subsection{Transformer-Based Autonomous Driving Perception Pipelines}

The adoption of transformer-based architectures in autonomous driving perception has accelerated significantly following the success of \ac{ViT}~\cite{dosovitskiy2021vit} and its hierarchical successors on standard vision benchmarks~\cite{liu2021swin}, alongside the broader real-time object-detection and hardware-acceleration literature for autonomous vehicles~\cite{sali2025realtimeodav}. The core appeal of transformers for autonomous driving lies in their ability to model long-range dependencies and global context across the full sensor field of view, properties critical when reasoning about spatially distant but semantically related objects. This has driven their adoption not only as image feature extractors but also as the backbone of unified perception, prediction, and planning frameworks~\cite{chen2025survey}.

Among hierarchical transformer backbones, the Swin Transformer~\cite{liu2021swin} has emerged as the de facto standard for multi-scale visual perception in autonomous driving, owing to its linear complexity, shifted-window cross-region interaction, and multi-resolution feature pyramid. UniAD~\cite{hu2023uniad} uses a Swin-T backbone to feed a cascade of task-specific transformer modules covering tracking, mapping, motion forecasting, occupancy prediction, and ego-motion planning, demonstrating that a single hierarchical encoder can simultaneously serve multiple heterogeneous perception tasks. SparseDrive~\cite{sun2024sparsedrive} similarly adopts a Swin backbone within a sparse instance query framework, exploiting its multi-resolution feature pyramid for efficient scene representation. The recurrence of hierarchical vision transformers across these systems underscores that their efficient deployment on embedded \ac{AI-SoC} hardware, specifically with full \ac{DLA} compatibility, is a critical prerequisite for practical autonomous driving, which is precisely the problem this work addresses.
\vspace{-1em}
\subsection{Edge-Aware Implementations of Transformer Models}

Prior work has explored the deployment of vision transformer models across a range of edge hardware platforms, employing strategies such as operator approximation, mixed-precision quantization, and hardware-software co-design to bridge the gap between transformer architecture requirements and the constraints of fixed-function accelerators \cite{liu2023efficientfpgabasedacceleratorswin, fu2024softact}, with recent FPGA-focused surveys extending this discussion to transformers, VLMs, CNN-based detection, classification, and tracking~\cite{sali2025fpgatransformersvlm, sali2025fpgacnns}. A notable example targeting the Swin Transformer specifically is \cite{liu2023efficientfpgabasedacceleratorswin}, which proposes an \ac{FPGA}-based accelerator for the Swin Transformer targeting edge computing applications such as autonomous driving and face recognition, where real-time inference is critical. The primary challenge lies in adapting computationally expensive operations, namely layer normalization, softmax, and \acf{GELU}, to fixed-point \ac{FPGA} logic. To address this, layer normalization is replaced with batch normalization, which can be used with preceding linear layers at inference time, eliminating the need for per-sample mean and variance computation. The softmax function is approximated by substituting the natural exponential $e^x$ with a base-2 power $2^x$, yielding:
\begin{equation}
    \text{Softmax}(x_i) \approx \frac{2^{x_i - x_{\max}}}{\sum_{j} 2^{x_j - x_{\max}}}
    \label{eq:softmax_approx}
\end{equation}
\noindent This reformulation reduces exponentiation to hardware-efficient bit-shift and addition operations. The \ac{GELU} activation is similarly approximated using a sigmoid-based form:
\begin{equation}
    \text{GELU}(x) \approx x \cdot \sigma(1.702\,x)
    \label{eq:gelu_approx}
\end{equation}
\noindent where $\sigma(\cdot)$ denotes the sigmoid function, avoiding the need for hyperbolic tangent or Gaussian cumulative distribution function evaluation. All linear operations, including convolutions and matrix multiplications, are unified by a single matrix-multiplication unit operating on a 16-bit fixed-point datapath. Against a CPU baseline (AMD Ryzen 5700X), the accelerator achieves a $1.25\times$ speedup on Swin-B (base model of Swin Transformer), with energy efficiency gains of $14.63\times$ over the CPU and $3.00\times$ over an NVIDIA RTX 2080 Ti GPU, at the cost of lower raw throughput ($0.12\times$ that of the GPU). The substitution of layer normalization with batch normalization incurs a $0.7\%$ accuracy loss on Swin-B, which the authors consider acceptable given the efficiency gains~\cite{liu2023efficientfpgabasedacceleratorswin}.

Map-and-Conquer \cite{bouzidi2023mapandconquer} takes a complementary approach by partitioning AI models horizontally along the width dimension into concurrent sub-networks, which are then dispatched across heterogeneous compute units such as GPU and \ac{DLA} cores. Applied to the Visformer vision transformer \cite{visformer} on the NVIDIA Jetson AGX Xavier, the scheduler reduces energy consumption by $14\%$ and latency by $43\%$ relative to \ac{DLA}-only execution, while preserving model accuracy. However, Map-and-Conquer assumes full layer compatibility with the target hardware and does not handle \ac{DLA}-incompatible layers. This limits its applicability to more complex architectures such as the Swin Transformer, where certain layers cannot be efficiently mapped to \ac{DLA} cores without additional adaptation.

\section{Background}

\subsection{Hardware Architecture}
The NVIDIA Jetson AGX Orin was considered in this paper as it incorporates the same fundamental heterogeneous computing components found in many modern edge AI platforms, including 
multi-core ARM CPUs, a high-performance GPU, and dedicated accelerators, all sharing a unified memory subsystem. The Orin device combines multi-core ARM CPUs, an Ampere-based GPU, dual \ac{DLA} engines, and additional multimedia accelerators such as the \ac{PVA}, \ac{VIC}, and \ac{OFA}, all interconnected through a shared memory subsystem.  

\subsubsection{\textbf{GPU}} The Orin GPU is an Ampere-architecture device with two \acp{GPC}, each containing 8 \acp{SM} with 128 CUDA cores and 64 Tensor Cores, delivering up to 170 Sparse TOPS \cite{xavier-orin-data}.

\subsubsection{\textbf{\ac{DLA}}} The \ac{DLA} is a fixed-function accelerator designed for energy-efficient CNN inference, comprising a convolution core, data processors, dedicated memory, and reshape engines \cite{nvdla-doc}. Its integration within an edge GPU platform is architecturally significant. By executing structured, computationally intensive operations such as convolutions and pooling, the \ac{DLA} frees GPU resources for more complex, irregular computations, thereby enabling heterogeneous parallel execution. This capability is particularly important in autonomous driving and other edge-AI applications, where multiple perception and decision-making tasks must execute concurrently under stringent power and thermal constraints. Furthermore, the \ac{DLA} typically delivers significantly higher performance per watt than the GPU for convolution-dominated workloads, making it an attractive execution target for compatible portions of deep neural networks. Consequently, there has been growing interest in scheduling and partitioning AI workloads across the \ac{DLA} and GPU to maximize system throughput and energy efficiency \cite{review}. 

However, several hardware constraints limit the direct deployment of transformer-based models on the \ac{DLA} \cite{dla-limit}. Supported precisions are restricted to FP16 and INT8, tensors must be four-dimensional, activation functions are limited to ReLU, Sigmoid, TanH, Clipped ReLU, and Leaky ReLU, and neither \ac{GELU} activations nor large matrix multiplications are natively supported. Operators that violate these constraints are automatically offloaded to the GPU, resulting in execution fragmentation, reduced concurrency, and lower overall throughput. The term DLA-targeted execution is used throughout this paper to indicate TensorRT deployment with the \ac{DLA} selected as the primary execution engine. Unsupported operations may still execute via TensorRT-managed GPU fallback; therefore, the objective of the proposed adaptation is to reduce, rather than eliminate, fallback-induced fragmentation.

\definecolor{IEEEpurple}{HTML}{981D97}   
\definecolor{IEEEdpurple}{HTML}{772583}
\definecolor{IEEEcyan}{HTML}{00B5E2}     
\definecolor{IEEEblue}{HTML}{00629B}
\definecolor{IEEEteal}{HTML}{009CA6}     
\definecolor{IEEEdteal}{HTML}{007377}
\definecolor{IEEEgreen}{HTML}{78BE20}    
\definecolor{IEEEolive}{HTML}{658D1B}
\definecolor{IEEEgray}{HTML}{75787B}     
\definecolor{IEEElgray}{HTML}{F2F2F2}
\definecolor{IEEEred}{HTML}{BA0C2F}

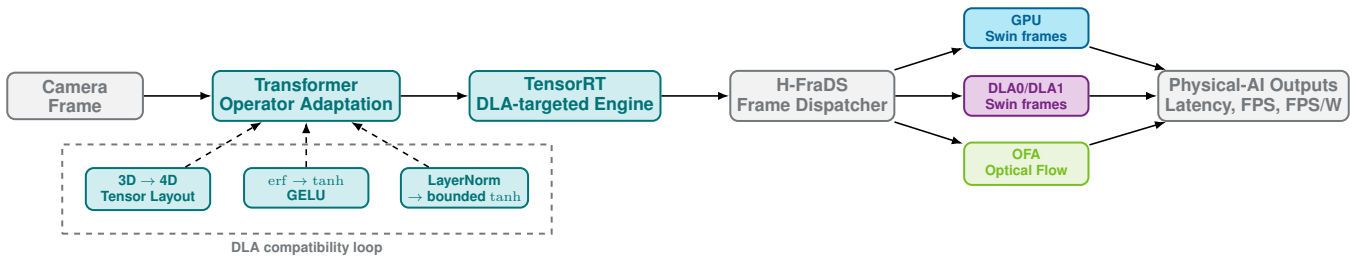
\begin{figure*}[!b]
    \centering
    \resizebox{0.98\linewidth}{!}{%
    \begin{tikzpicture}[
        node distance=8mm and 12mm,
        block/.style={rounded corners, align=center, minimum height=8mm,
            minimum width=24mm, font=\sffamily\small\bfseries, line width=1pt},
        smallblock/.style={rounded corners, align=center, minimum height=7mm,
            minimum width=22mm, font=\sffamily\scriptsize\bfseries, line width=1pt},
        arrow/.style={-{Latex[length=2mm]}, thick, draw=black}
    ]

    \node[block, fill=IEEElgray, draw=IEEEgray, text=IEEEgray] (cam) {Camera\\Frame};

    \node[block, right=of cam, fill=IEEEteal!18, draw=IEEEdteal, text=IEEEdteal]
        (adapt) {Transformer\\Operator Adaptation};

    \node[smallblock, below=of adapt, xshift=-28mm, fill=IEEEteal!18, draw=IEEEdteal, text=IEEEdteal] (reshape) {3D $\rightarrow$ 4D\\Tensor Layout};
    \node[smallblock, below=of adapt, fill=IEEEteal!18, draw=IEEEdteal, text=IEEEdteal]
        (gelu) {$\mathrm{erf}\rightarrow\tanh$\\GELU};
    \node[smallblock, below=of adapt, xshift=28mm, fill=IEEEteal!18, draw=IEEEdteal, text=IEEEdteal] (ln) {LayerNorm\\$\rightarrow$ bounded $\tanh$};

    \node[block, right=of adapt, fill=IEEEteal!18, draw=IEEEdteal, text=IEEEdteal]
        (trt) {TensorRT\\DLA-targeted Engine};

    \node[block, right=of trt, fill=IEEElgray, draw=IEEEgray, text=IEEEgray]
        (dispatch) {\hfrads\\Frame Dispatcher};

    \node[smallblock, right=of dispatch, yshift=12mm, fill=IEEEcyan!30, draw=IEEEblue,
        text=IEEEblue] (gpu) {GPU\\Swin frames};
    \node[smallblock, right=of dispatch, fill=IEEEpurple!22, draw=IEEEdpurple,
        text=IEEEdpurple] (dla) {DLA0/DLA1\\Swin frames};
    \node[smallblock, right=of dispatch, yshift=-12mm, fill=IEEEgreen!15, draw=IEEEgreen,
        text=IEEEgreen] (ofa) {OFA\\Optical Flow};

    \node[block, right=of dla, fill=IEEElgray, draw=IEEEgray, text=IEEEgray]
        (metrics) {Physical-AI Outputs\\Latency, FPS, FPS/W};

    \draw[arrow] (cam) -- (adapt);
    \draw[arrow] (adapt) -- (trt);
    \draw[arrow] (trt) -- (dispatch);
    \draw[arrow] (dispatch) -- (gpu);
    \draw[arrow] (dispatch) -- (dla);
    \draw[arrow] (dispatch) -- (ofa);
    \draw[arrow] (gpu) -- (metrics);
    \draw[arrow] (dla) -- (metrics);
    \draw[arrow] (ofa) -- (metrics);

    \draw[arrow, dashed] (reshape) -- (adapt);
    \draw[arrow, dashed] (gelu) -- (adapt);
    \draw[arrow, dashed] (ln) -- (adapt);

    \node[draw=IEEEgray, dashed, line width=1pt, fit=(reshape)(gelu)(ln),
        inner sep=4mm,
        label=below:{\sffamily\scriptsize\bfseries\color{IEEEgray} DLA compatibility loop}] {};

    \end{tikzpicture}}
    \caption{\small Proposed edge physical AI deployment pipeline. A hardware-aware Swin Transformer processes incoming camera frames, compiled using TensorRT and executed through the H-FraDS dispatcher, which routes frames across GPU, DLA0, DLA1, and OFA pipelines according to the selected dispatch ratio. }    
    \label{fig:native_system_overview}
\end{figure*}

\subsubsection{\textbf{\ac{OFA}}} The \ac{OFA} is a fixed-function engine for optical flow and stereo disparity \cite{xavier-orin-data}. In this work, optical flow mode is used: the \ac{OFA} accepts an externally built image pyramid (scale factor 2, up to five levels) and outputs per-block S10.5 motion vectors at grid sizes of $1{\times}1$ to $8{\times}8$ pixels, achieving 1.13~ms latency at $1920{\times}1080$ with 8${\times}$8 grid size \cite{vpi}.

\vspace{-1em}
\subsection{Transformer Architecture}
The transformer architecture was originally introduced by Vaswani et al.~\cite{vaswani2017attention} for sequence-to-sequence tasks in natural language processing. Its core building block is the self-attention mechanism, which computes pairwise relationships between all elements in a sequence. Given an input sequence of $n$ tokens, each represented as a $d$-dimensional vector, the mechanism projects the input into queries $\mathbf{Q}$, keys $\mathbf{K}$, and values $\mathbf{V}$ via learned linear projections, and computes attention as:
{\setlength{\abovedisplayskip}{1pt}
\setlength{\belowdisplayskip}{1pt}
\begin{equation}
    \text{Attention}(\mathbf{Q}, \mathbf{K}, \mathbf{V}) = \text{Softmax}\!\left(\frac{\mathbf{Q}\mathbf{K}^\top}{\sqrt{d_k}}\right)\mathbf{V}
    \label{eq:attention}
\end{equation}}

\noindent where $d_k$ is the dimension of the keys. Multiple attention heads are computed in parallel and concatenated (Multi-head Self-Attention, MSA), enabling the model to capture relationships at different representation subspaces simultaneously~\cite{vaswani2017attention}. Each transformer block pairs MSA with a two-layer \ac{MLP} using \ac{GELU} activation~\cite{GELU}, with layer normalization~\cite{layer_norm} applied before each sub-layer following the pre-norm convention.

The \ac{ViT}~\cite{dosovitskiy2021vit} adapted this architecture to image data by splitting the input image into a sequence of fixed-size non-overlapping patches, linearly projecting each patch into a token embedding, and appending a learnable classification token. A stack of standard transformer encoder blocks processes the resulting sequence. While \ac{ViT} demonstrated competitive accuracy on large-scale image classification, its global self-attention over a fixed-resolution sequence produces a single-scale feature map and scales quadratically with image resolution, both of which are significant limitations for dense prediction tasks and high-resolution automotive cameras.

Hierarchical vision transformers address these limitations by introducing multi-scale feature extraction and local attention. Rather than computing attention globally, attention is restricted to local spatial windows, reducing the complexity from quadratic to linear with respect to image size. Cross-window interactions are recovered through a shifted-window mechanism that alternates the window partition between successive blocks~\cite{liu2021swin}. Patch merging layers downsample the spatial resolution between stages while increasing channel depth, producing a multi-scale feature pyramid analogous to a CNN backbone and well-suited to object detection and semantic segmentation. This design has become the dominant backbone for autonomous driving perception frameworks~\cite{hu2023uniad, sun2024sparsedrive, chen2025survey, LiuZhengyi2022SSTD}.

Three structural properties are shared across this family of architectures and are directly responsible for \ac{DLA} incompatibility. First, the self-attention module naturally produces intermediate tensors of the form $(\text{Batch} \times \text{Sequence} \times \text{Channel})$, a 3D layout incompatible with the \ac{DLA}'s fixed 4D addressing scheme. Second, the \ac{GELU} activation~\cite{GELU} relies on the \acf{ERF} function, which is not a natively supported primitive on the \ac{DLA}. Third, layer normalization~\cite{layer_norm} requires dynamic per-sample computation of mean and variance, operations that the \ac{DLA} cannot execute. Together, these three incompatibilities force GPU fallback at multiple points in the pipeline, fragmenting execution and reducing the energy efficiency benefit of the accelerator. The methodology proposed in Section~\ref{sec:methodology} addresses each incompatibility in turn and is demonstrated on the Swin Transformer~\cite{liu2021swin}, a representative hierarchical vision transformer that exhibits all three.

\section{Suggested Methodology}
\label{sec:methodology}

\subsection{Hardware Accelerator-Aware Model}
Fig.~\ref{fig:native_system_overview} summarizes the proposed native system view, replacing generic architecture diagrams with the complete edge physical AI deployment flow from camera input to operator adaptation, TensorRT compilation, heterogeneous frame dispatch, and platform-level efficiency measurement. To produce a \ac{DLA}-compatible variant of a vision transformer, each model stage must be systematically analyzed to identify operations and tensor formats incompatible with the \ac{DLA}'s hardware constraints. For each identified incompatibility, a functionally equivalent or mathematically approximated substitute is applied to ensure that the modified model can execute as a continuous \ac{DLA} pipeline with minimal GPU fallback. This methodology is extended here and applied to transformer architectures, following the broader hardware-aware deployment philosophy of prior convolutional-model implementations~\cite{mri_pipeline}. The challenge of hardware-efficient execution of transformer operations, including large matrix multiplications, nonlinear activations, and normalization layers, has been studied from the FPGA perspective in \cite{10623817, fu2024softact, liu2023efficientfpgabasedacceleratorswin}. This work addresses the similar incompatibilities from the \ac{DLA} perspective on an edge \ac{AI-SoC}. The following subsections detail each incompatibility class identified in the Swin Transformer, together with its corresponding substitution. The same substitutions apply to any transformer architecture exhibiting these structural patterns.

\subsubsection{\textbf{3D Format}} 
The \ac{DLA} requires all input and intermediate tensors to conform to a 4D format, represented as either $(\text{Batch} \times \text{Channel} \times \text{Height} \times \text{Width})$ or $(\text{Batch} \times \text{Height} \times \text{Width} \times \text{Channel})$ depending on the data layout convention \cite{dla-limit}. This is a fundamental hardware limitation, as the \ac{DLA}'s memory-addressing circuitry is physically designed for 4D spatial feature maps and lacks an addressing mode for 3D tensors \cite{nvdla-doc}. It cannot be resolved through software or driver updates; the memory controller does not support the addressing pattern required by a 3D tensor. Within the transformer's self-attention module, intermediate tensors of the form $(\text{Batch} \times \text{Sequence} \times \text{Channel})$ are naturally produced \cite{liu2021swin} and are therefore structurally incompatible with this hardware requirement. Without modification, these operations fall back to the GPU, necessitating the insertion of additional shuffle layers at the \ac{DLA}-GPU boundaries to resize tensors between the two formats for each transition.

To eliminate this overhead, all 3D tensors were reshaped into their 4D equivalents by reintroducing the spatial dimensions prior to \ac{DLA} execution. This ensures that tensor operations remain within the supported format throughout inference, removing the need for GPU fallback at these layers. As a direct consequence, the number of shuffle layers required across the pipeline was reduced, resulting in more continuous \ac{DLA} execution and improved overall inference efficiency.

\begin{figure}[ht]
    \centering
    \includegraphics[width=\linewidth]{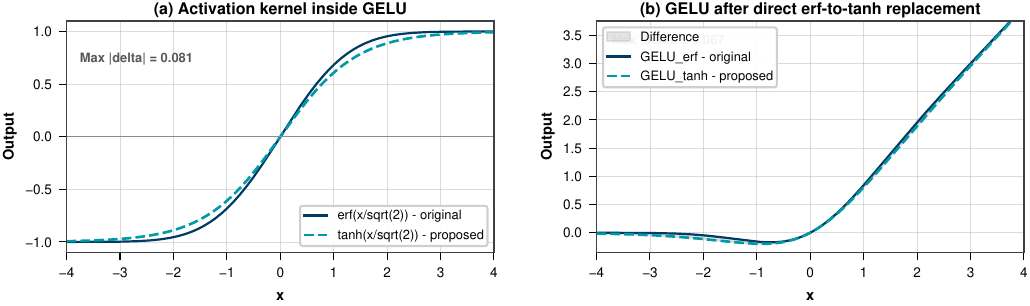}
    \caption{\small Left: comparison of $\mathrm{erf}(x)$ and $\tanh(x)$, illustrating their close agreement. Right: the resulting \ac{GELU} activation computed with each formulation, showing a negligible difference between the erf-based original and the $\tanh$-based approximation used in the proposed model.}
    \label{fig:gelu_comparison}
\end{figure}

\subsubsection{\textbf{\ac{GELU}}} 
The GELU activation function can be expressed using Equation \ref{eq:gelu} \cite{GELU}. The root cause of its \ac{DLA} incompatibility is the presence of the \ac{ERF} function, which is absent from the \ac{DLA}'s natively supported primitive set \cite{dla-limit}. This is primarily a hardware limitation, as the \ac{DLA} implements activation functions as fixed lookup tables or dedicated arithmetic units, and the \ac{ERF} function, which requires evaluating a Gaussian integral, is not among the operations supported by the \ac{DLA}'s hardware. This is not a software stack deficiency that could be resolved with a driver or compiler update; the physical compute units needed to evaluate \ac{ERF} are not present. As shown in Equation \ref{eq:erf}, the \ac{ERF} function can be closely approximated using $\tanh$ \cite{GELU}.

\begin{align} \label{eq:gelu}
    \text{GELU}(x) = \frac{x}{2}\left[1 + \text{erf}\!\left(\frac{x}{\sqrt{2}}\right)\right] 
\end{align}

{\setlength{\abovedisplayskip}{0.5pt}
\setlength{\belowdisplayskip}{0.5pt}
\begin{align} \label{eq:erf}
    \text{erf}(x) = \frac{2}{\sqrt{\pi}} \int_0^x e^{-t^2} dt \approx \tanh\!\left(\sqrt{\frac{2}{\pi}}\left(x + 0.044715x^3\right)\right) 
\end{align}
}

In our implementation, the \ac{ERF} within the \ac{GELU} activation was directly replaced with the $\tanh$ function for simplicity, as the $\tanh$ is natively supported by the \ac{DLA} as a standard activation function \cite{dla-limit}. This targeted substitution preserves the overall structure of \ac{GELU} while eliminating the unsupported \ac{ERF} operation, allowing the activation function to remain \ac{DLA}-targeted with reduced fallback. The resulting approximated \ac{GELU} is given in Equation \ref{eq:gelu_new}.

{\setlength{\abovedisplayskip}{0.20pt}
\begin{align} \label{eq:gelu_new}
    \text{GELU}(x) = \frac{x}{2}\left[1 + \tanh\!\left(\frac{x}{\sqrt{2}}\right)\right] 
\end{align}
}

The \ac{GELU} substitution is motivated by the standard tanh-form approximation of the Gaussian cumulative distribution used in transformer implementations. Unlike the exact \ac{ERF}-based expression, the tanh-based form maps directly to a \ac{DLA}-supported activation primitive. The approximation preserves the smooth gating behavior of \ac{GELU} while avoiding unsupported transcendental operations.

\begin{figure}[h]
    \centering
    \includegraphics[width=\linewidth]{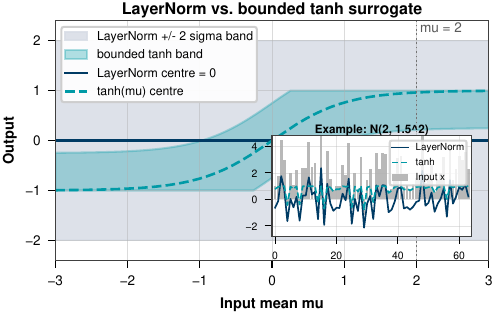}
    \caption{\small Effect of layer normalization and the $\tanh$ approximation on an example feature vector drawn from $\mathcal{N}(2,\,1.5^2)$ (inset), and output range as a function of the input mean $\mu$ (main).}
    \label{fig:layernorm_comparison}
\end{figure}

\subsubsection{\textbf{Layer Normalization}}
Layer normalization, introduced by \cite{layer_norm}, is applied throughout the Swin Transformer after each attention block and \ac{MLP} block \cite{liu2021swin}. Equation \ref{eq:layer_norm} represents the operation, where $\mu$ and $\sigma^2$ are the mean and variance computed across the feature dimension of the input, $\epsilon$ is a small constant added for numerical stability, and $\gamma$ and $\beta$ are learned affine parameters \cite{layer_norm}. 
\vspace{-0.5em}
\begin{align} \label{eq:layer_norm}
    \text{Layer normalization}(x) = \frac{x - \mu}{\sqrt{\sigma^2 + \epsilon}} \cdot \gamma + \beta
\end{align}

The \ac{DLA} does not support the dynamic computation of mean and variance, nor the square root and division operations required by layer normalization \cite{dla-limit}. This is again a hardware limitation as these operations require data-dependent, per-sample arithmetic that the \ac{DLA}'s fixed-function pipeline cannot perform — unlike a GPU, which executes arbitrary compute shaders, the \ac{DLA} has no programmable execution units capable of computing running statistics. This cannot be resolved at the software or compiler level, as no amount of graph optimization can map dynamic normalization onto hardware that physically lacks the required arithmetic capability. This forces these layers to be executed on the GPU, resulting in pipeline fragmentation. To improve \ac{DLA}-targeted execution, layer normalization was substituted with a $\tanh$-based approximation, which is natively supported. While $\tanh$ does not replicate the exact normalization behavior of layer normalization \cite{layer_norm}, it provides a bounded, smooth non-linearity that can partially compensate for the distribution shift that layer normalization would otherwise correct. 

\begin{figure*}[!b]
    \centering
    \subfloat[Concurrent Task Scheduling. YOLOv8 runs on the GPU and Swin Transformer on the \ac{DLA} in free-running, independent threads with no shared dispatch clock.\label{fig:timing_naive}]{%
        \includegraphics[width=0.48\linewidth]{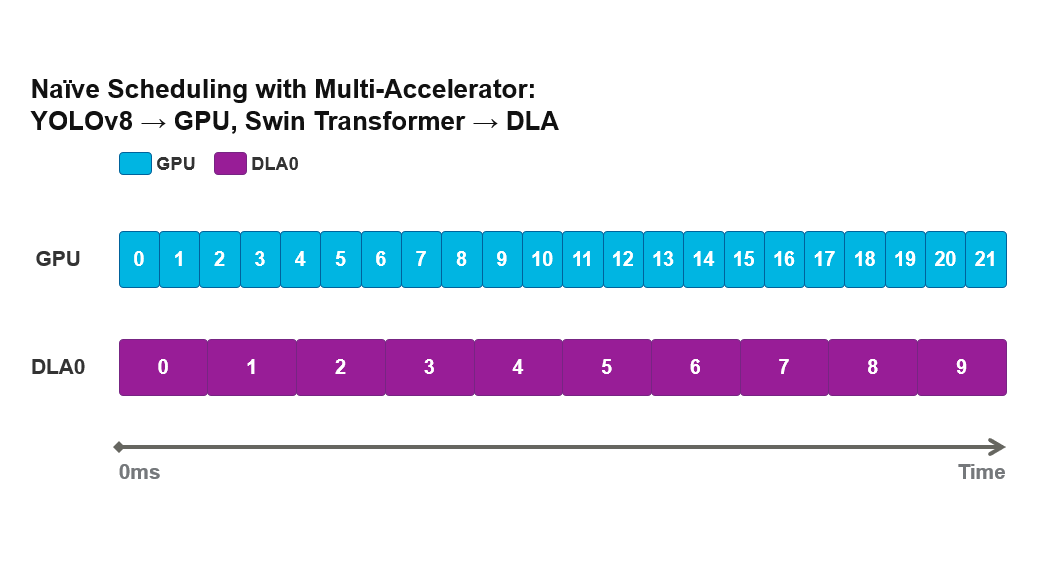}}\hfill
    \subfloat[Multi-Accelerator Scheduling. YOLOv8, Swin Transformer, and Dense Optical Flow run concurrently on the GPU, \ac{DLA}, and \ac{OFA}, respectively, in independent free-running threads.\label{fig:timing_ofa}]{%
        \includegraphics[width=0.48\linewidth]{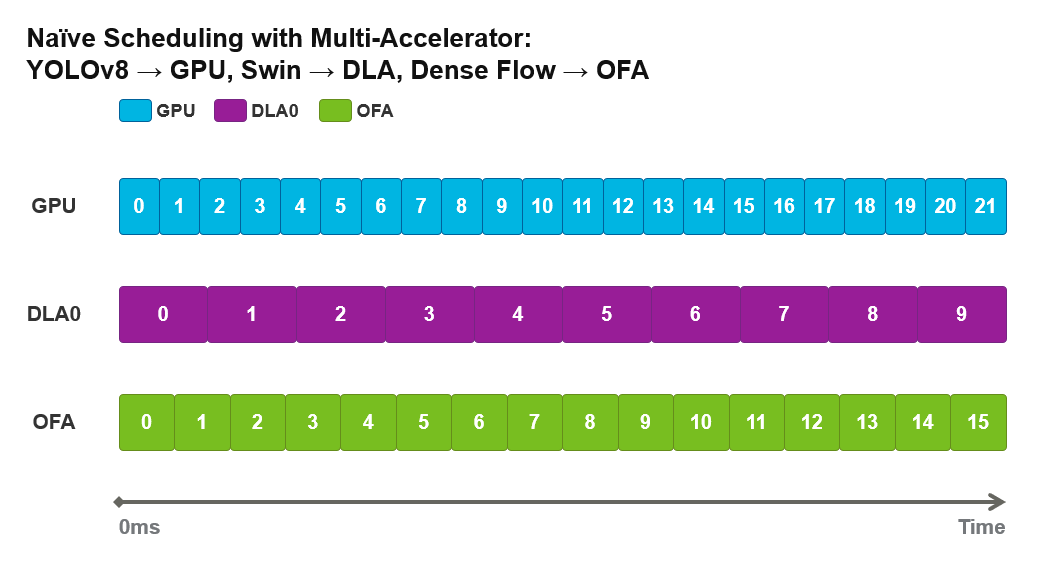}}
    \caption{ \small Timing diagrams for free-running concurrent scheduling strategies.}
    \label{fig:timing_naive_ofa}
\end{figure*}

The LayerNorm substitution is not an algebraic equivalence. Instead, it is a deployment-oriented bounded surrogate introduced to preserve stable activation ranges under \ac{DLA} constraints. LayerNorm performs three functions: centering, variance normalization, and learned affine rescaling. The proposed bounded $\tanh$ surrogate primarily preserves the range-stabilization function, while sacrificing exact per-sample centering and variance normalization. This design choice is justified only when the resulting accuracy degradation remains acceptable after calibration or fine-tuning, as verified experimentally by the F1-score comparison. To quantify the approximation effect, operator-level activation deviation can be measured as
\begin{equation}
    \Delta_{\mathrm{act}} = \frac{1}{M}\sum_{i=1}^{M}\left|y_i^{\mathrm{orig}} - y_i^{\mathrm{sub}}\right|.
    \label{eq:activation_deviation}
\end{equation}
This metric separates operator-level approximation error from end-task accuracy degradation and provides a direct way to analyze the effect of the substitution. In this work, the end-to-end F1 degradation remains limited to 2.65 percentage points, suggesting that the substituted model preserves most task-relevant representations.

\begin{table}[H]
\centering
\caption{\small Theoretical comparison between LayerNorm and the bounded $\tanh$ surrogate.}
\label{tab:ln_tanh_theory}

\renewcommand{\arraystretch}{0.8}
\setlength{\tabcolsep}{2.2pt}

\begin{tabularx}{\columnwidth}{@{}
>{\raggedright\arraybackslash}p{0.23\columnwidth}
>{\raggedright\arraybackslash}X
>{\raggedright\arraybackslash}X
@{}}
\toprule
\textbf{Property} &
\textbf{LayerNorm} &
\textbf{Bounded $\tanh$ surrogate} \\
\midrule

Mean handling &
Explicit per-sample centering &
No explicit centering \\

Variance handling &
Explicit per-sample scaling &
Range compression \\

Output range &
Unbounded after affine parameters &
Bounded to $[-1,1]$ before later scaling \\

\ac{DLA} support &
Unsupported dynamic statistics, division, and square root &
Supported activation primitive \\

Interpretation &
Statistical normalization &
Hardware-feasible activation stabilization \\

\bottomrule
\end{tabularx}

\vspace{-0.8em}
\end{table}

For clarity, Table~\ref{tab:ln_tanh_theory} summarizes the theoretical difference between exact LayerNorm and the proposed bounded surrogate. The surrogate should be interpreted as a hardware-feasible stability mechanism rather than as a replacement with identical statistics. LayerNorm maps each token feature vector to a normalized affine space using data-dependent mean and variance. The bounded $\tanh$ surrogate instead compresses activations into $[-1,1]$, limiting outliers and reducing dynamic-range variation before \ac{DLA} execution. This is useful for fixed-function inference because bounded activations reduce the probability of large intermediate values that amplify quantization and scheduling sensitivity.

Consequently, the validity of the substitution is established empirically rather than by algebraic equivalence. The proposed model is acceptable only if the task-level accuracy drop remains bounded and the deployment benefit is significant. This is why the paper reports both the F1-score reduction and the throughput/latency/FPS-per-watt improvement, and why an operator-ablation table is included to isolate the contribution of each transformation.
\vspace{-1em}
\subsection{DLA-Aware Scheduling Strategies}
\label{sec:scheduling}
 
After compatibility adaptation, the scheduling stage evaluates how the available hardware engines can be used concurrently rather than treating the \ac{DLA} as an isolated execution target. Deploying the \ac{DLA}-compatible Swin Transformer on the Jetson AGX Orin in isolation exploits only one of the platform's inference-capable accelerators. To fully leverage the heterogeneous compute fabric, GPU, dual \ac{DLA} cores, and \ac{OFA}, this work designs and evaluates five scheduling strategies that distribute incoming perception frames or concurrent workloads across these units. The strategies are listed as follows: \emph{free-running concurrent} execution, \emph{free-running multi-accelerator} execution, \emph{free-running dual-\ac{DLA} symmetric execution}, single-\ac{DLA} \emph{\hfrads}, and dual-\ac{DLA} interleaved \emph{\hfrads}. In the timing diagram, the hatched regions indicate periods during which the hardware unit is idle, waiting for the next frame to arrive. In free-running strategies (Naive, Dual DLA, \ac{OFA}), no frame pacing is applied, and units loop at the hardware's fastest rate.

\subsubsection{\textbf{Free-Running Concurrent and Multi-Accelerator Scheduling}}
 
In the simplest configuration, separate inference tasks are assigned to individual accelerators and executed in independent threads with no shared dispatch clock. A YOLOv8 model runs on the GPU while the Swin Transformer runs on a single \ac{DLA} core, each looping at its own hardware-limited rate. An extended variant adds a dense optical flow model on the \ac{OFA}, creating a three-way concurrent pipeline. These strategies maximize individual accelerator utilization but do not coordinate frame arrival, and the \ac{DLA}'s throughput is limited by the per-inference latency of the transformer model. The timing diagrams for these free-running strategies are shown in Fig.~\ref{fig:timing_naive_ofa}. As neither the GPU nor the DLA units are paced, both loops continuously: the GPU completes YOLOv8 inference in $\approx$4 ms cycles, while the DLA processes Swin Transformer frames at $\approx$20 ms intervals in the proposed model. The absence of coordination means that any latency improvement from reducing DLA inference time directly translates into a proportional throughput gain, without introducing idle periods.

\begin{figure}[ht]
    \centering
    \includegraphics[width=\linewidth]{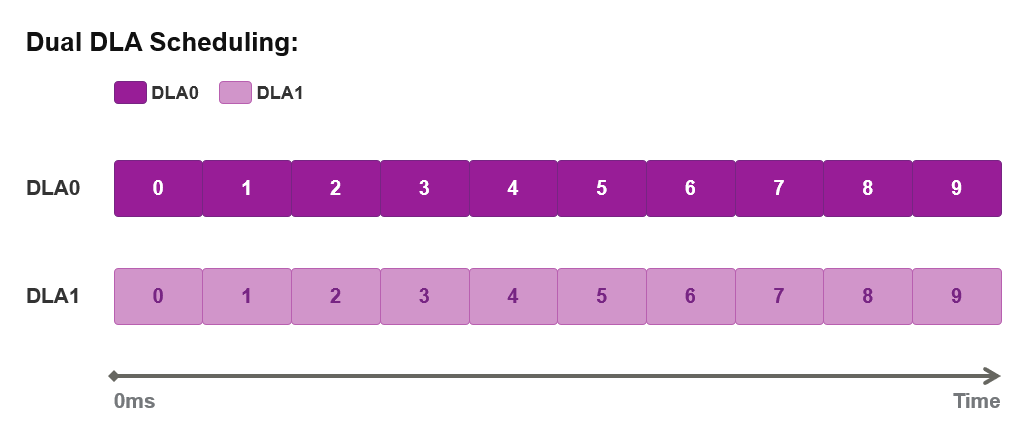}
    \caption{ \small Timing diagram for Dual DLA scheduling. Two Swin Transformer instances run on \ac{DLA} core 0 and \ac{DLA} core 1, respectively, in independent free-running threads.}
    \label{fig:timing_dual_dla}
\end{figure}

\subsubsection{\textbf{Free-Running Symmetric Dual-\ac{DLA}  Scheduling}}
 
A symmetric dual-\ac{DLA} configuration runs two independent Swin Transformer instances on DLA0 and DLA1 in parallel, doubling raw \ac{DLA} throughput without GPU involvement. Because no explicit GPU inference task is launched, any observed GPU utilization in this mode is entirely due to GPU-fallback operations within the TensorRT \ac{DLA} engines. The timing diagram for the Dual-\ac{DLA} configuration is shown in Fig.~\ref{fig:timing_dual_dla}. Both \ac{DLA} cores execute independently and symmetrically, with no inter-core synchronization. Since no GPU inference task is scheduled, the GPU remains idle except for \ac{DLA}-fallback operations. The diagram illustrates that the two engines operate in near lock-step, confirming the symmetric hardware behavior observed in the results.

\definecolor{IEEEteal}{HTML}{009CA6}    
\definecolor{IEEEdteal}{HTML}{007377}
\definecolor{IEEEDarkGray}{HTML}{75787B}
\definecolor{IEEELightGray}{HTML}{F2F2F2}
\definecolor{IEEELineGray}{HTML}{000000}

\begin{figure}[h]
    \centering
    \resizebox{\columnwidth}{!}{%
    \begin{tikzpicture}[
        terminal/.style={draw=IEEEDarkGray,fill=IEEELightGray,rounded corners=3mm,minimum width=30mm,minimum height=8mm,align=center,font=\sffamily\small\bfseries,text=IEEEDarkGray,line width=1.0pt},
        decision/.style={diamond,draw=IEEEdteal,fill=IEEEteal!18,aspect=2.2,minimum width=34mm,minimum height=17mm,inner sep=1pt,align=center,font=\sffamily\small\bfseries,text=IEEEdteal,line width=1.0pt},
        process/.style={draw=IEEEdteal,fill=IEEEteal!18,rounded corners=2mm,minimum width=26mm,minimum height=9mm,align=center,font=\sffamily\small\bfseries,text=IEEEdteal,line width=1.0pt},
        branchlabel/.style={font=\sffamily\small\bfseries,text=black,align=center},
        flowlabel/.style={font=\sffamily\small\bfseries,text=black,fill=white,inner sep=1.4pt},
        arrow/.style={-{Latex[length=2.5mm,width=1.9mm]},line width=1.1pt,draw=IEEELineGray},
        plainline/.style={line width=1.1pt,draw=IEEELineGray}
    ]
        \node[terminal] (arrive) at (0, 0){Frame arrives};
        \node[process,minimum width=50mm,minimum height=10mm] (mod) at (0, -1.45){Compute FrameID mod N\\[-1pt]N set by chosen scheduling ratio};
        \node[decision] (mode) at (0, -3.35){Single or Dual\\DLA mode?};
        \node[branchlabel] (singlelabel) at (-4.15, -2.55){Single-DLA Frame\\Dispatch Scheduling};
        \node[branchlabel] (duallabel) at (4.15, -2.55){Dual-DLA Interleaved\\Frame Dispatch};
        \node[decision] (singledec) at (-4.15, -5.15){FrameID\\mod N = 0?};
        \node[decision] (dualdec) at (4.15, -5.15){FrameID\\mod N = 0\\or 1?};
        \node[process] (dla0) at (-4.15, -6.90){DLA0};
        \node[process] (gpu) at (0, -6.55){GPU};
        \node[process] (dla01) at (4.15, -6.90){DLA0 or DLA1};
        \node[terminal] (out) at (0, -9.00){Detection output};
 
        \coordinate (leftNoTurn)  at (-0.55, -5.15);
        \coordinate (rightNoTurn) at ( 0.55, -5.15);
        \coordinate (outbarL) at (-4.15, -8.25);
        \coordinate (outbarR) at ( 4.15, -8.25);
        \coordinate (outbarC) at ( 0.00, -8.25);
 
        \draw[arrow] (arrive) -- (mod);
        \draw[arrow] (mod) -- (mode);
 
        \draw[arrow] (mode.west) -| (singledec.north);
        \draw[arrow] (mode.east) -| (dualdec.north);
 
        \draw[arrow] (singledec.south) -- (dla0.north);
        \draw[plainline] (singledec.east) -- (leftNoTurn) node[midway,above,flowlabel]{No};
        \draw[arrow] (leftNoTurn) -- ([xshift=-5.5mm]gpu.north);
 
        \draw[arrow] (dualdec.south) -- (dla01.north);
        \draw[plainline] (dualdec.west) -- (rightNoTurn) node[midway,above,flowlabel]{No};
        \draw[arrow] (rightNoTurn) -- ([xshift=5.5mm]gpu.north);
 
        \draw[arrow] (gpu.south) -- (out.north);
 
        \draw[plainline] (dla0.south) -- ++(0,-0.65) node[pos=0.55,right,flowlabel]{Yes} -- (outbarL) -- (outbarC);
        \draw[plainline] (dla01.south) -- ++(0,-0.65) node[pos=0.55,left,flowlabel]{Yes} -- (outbarR) -- (outbarC);
    \end{tikzpicture}}
    \caption{\small Flowchart of the \hfrads~frame dispatch strategy. }
    \label{fig:scheduling_flowchart}
\end{figure}
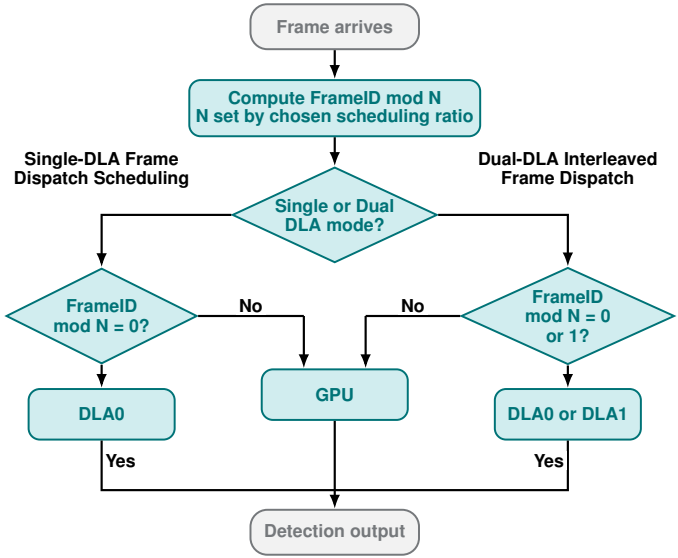

\begin{figure*}[!t]
    \centering
    \subfloat[1:1 (Alternating Frame). Both units finish well before the next frame arrives.\label{fig:even_odd}]{%
        \includegraphics[width=0.48\linewidth]{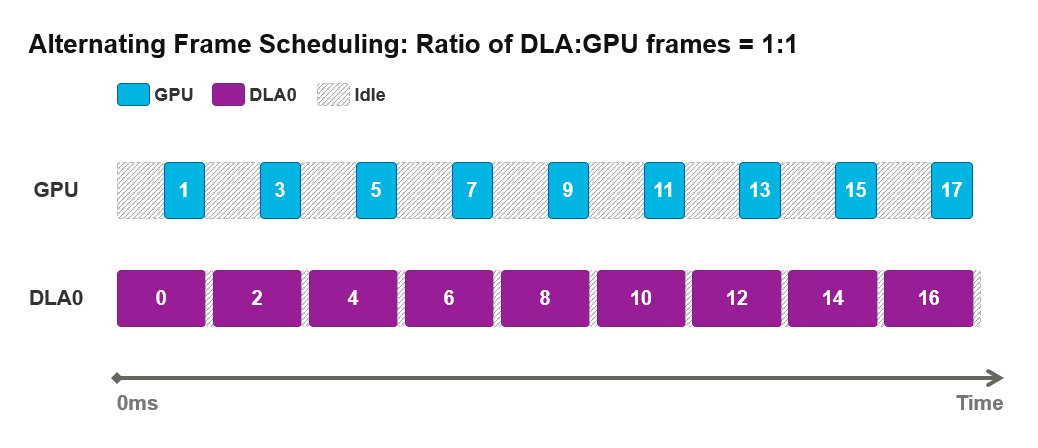}}\hfill
    \subfloat[1:2 (Balanced Dispatch). DLA latency ($\approx$20 ms) $\approx$ 2$\times$ GPU latency ($\approx$9 ms) — best-balanced ratio.\label{fig:timing_3rd}]{%
        \includegraphics[width=0.48\linewidth]{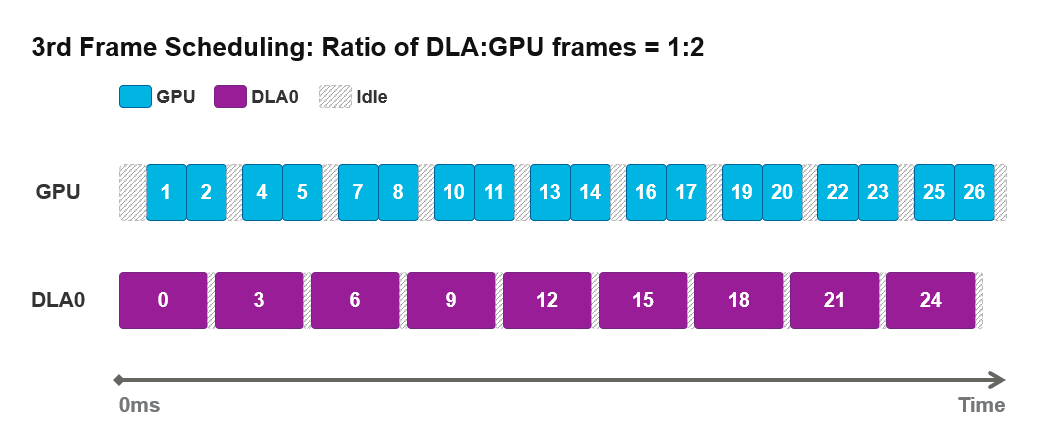}}\\[4pt]
    \subfloat[1:3 (GPU-Dominant). GPU becomes the bottleneck (3$\times$9\,ms $>$ 20\,ms), leaving the \ac{DLA} idle $\approx$7 ms per cycle.\label{fig:timing_4th}]{%
        \includegraphics[width=0.48\linewidth]{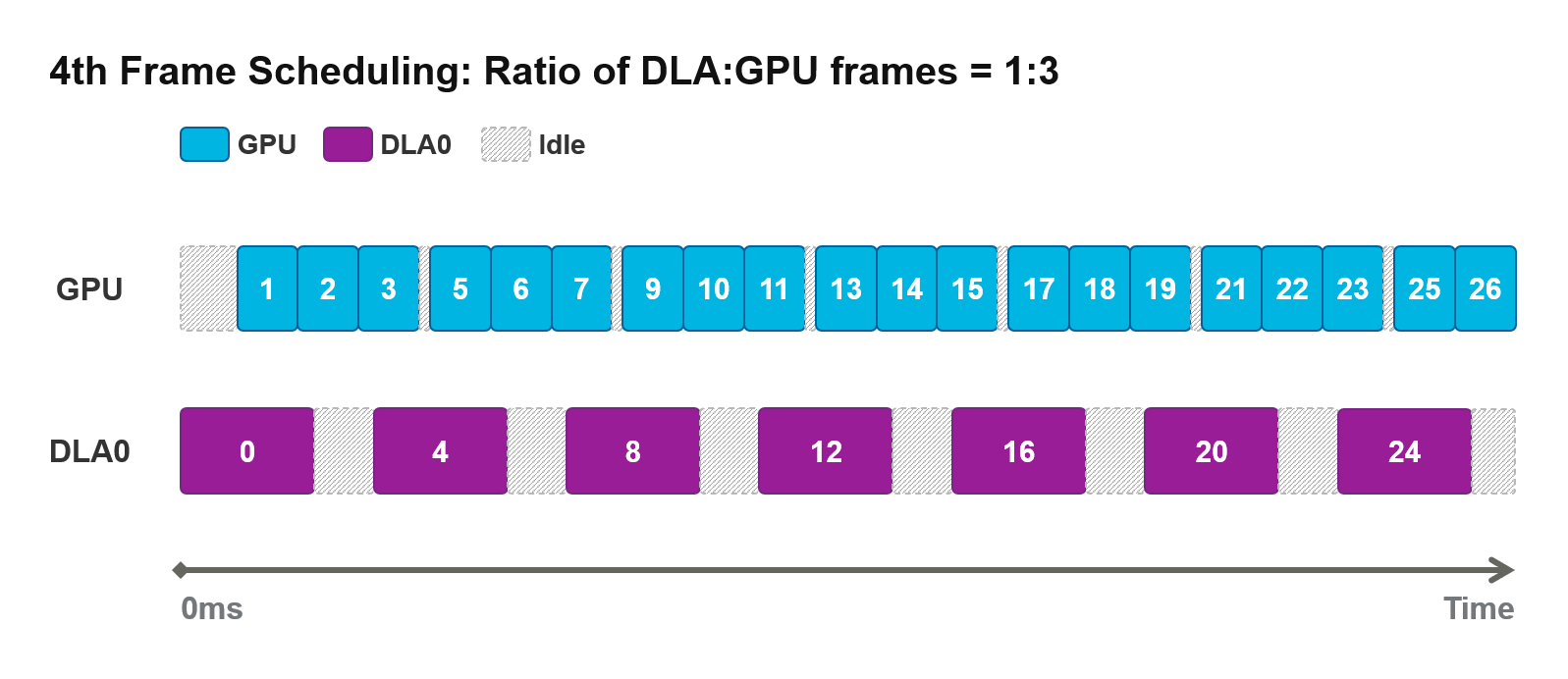}}\hfill
    \subfloat[1:5 (DLA-Sparse). \ac{DLA} finishes at $\approx$20 ms, but the next DLA frame does not arrive until 200 ms into the cycle.\label{fig:timing_6th}]{%
        \includegraphics[width=0.48\linewidth]{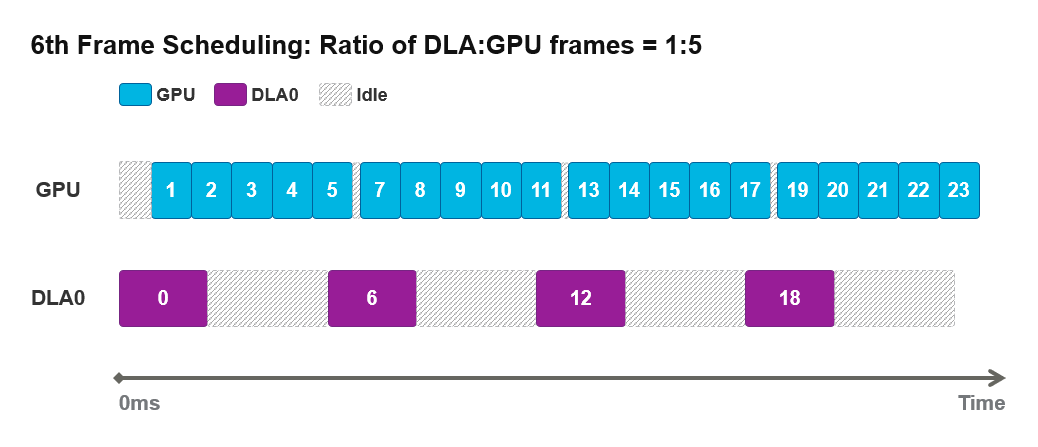}}
    \caption{ \small Timing diagrams for single-DLA \hfrads~strategies. Block widths are proportional to per-frame latencies; hatched regions indicate idle periods.}
    \label{fig:timing_frame_dispatch}
\end{figure*}

\subsubsection{\textbf{\hfrads: Single-\ac{DLA} and Dual-\ac{DLA} Frame Dispatch}} 
\hfrads~is the central contribution of this scheduling analysis. Rather than running each accelerator independently, a shared dispatch controller monitors incoming frames and routes them deterministically to either the \ac{DLA} or the GPU based on a fixed ratio $N$:$(N-1)$ (DLA:GPU). Every $N$-th frame in the sequence is forwarded to the \ac{DLA} core; the remaining $N-1$ frames per cycle are dispatched to the GPU. This design exploits the latency asymmetry between the two accelerators: the \ac{DLA} is slower per inference but energy-efficient, while the GPU is faster but power-hungry. By selecting a ratio that matches the hardware latency balance, both units can remain simultaneously busy with minimal idle time.

The flowchart in Fig.~\ref{fig:scheduling_flowchart} illustrates the \hfrads~routing logic. Four single-\ac{DLA} ratios are evaluated: 1:1 (Alternating Frame), 1:2 (Balanced Dispatch), 1:3 (GPU-Dominant), and 1:5 (DLA-Sparse). Two dual-\ac{DLA} variants extend this to route two consecutive frames, one to DLA0 and one to DLA1, per $N$-th cycle, at ratios 2:2 and 2:3.

\begin{figure*}[!t]
    \centering
    \subfloat[2:2 ratio. DLA0 receives $\text{frameId} \bmod 4 = 0$; DLA1 receives $\text{frameId} \bmod 4 = 1$; remaining frames go to the GPU. Both DLA cores run concurrently.\label{fig:timing_4th_dual}]{%
        \includegraphics[width=0.48\linewidth]{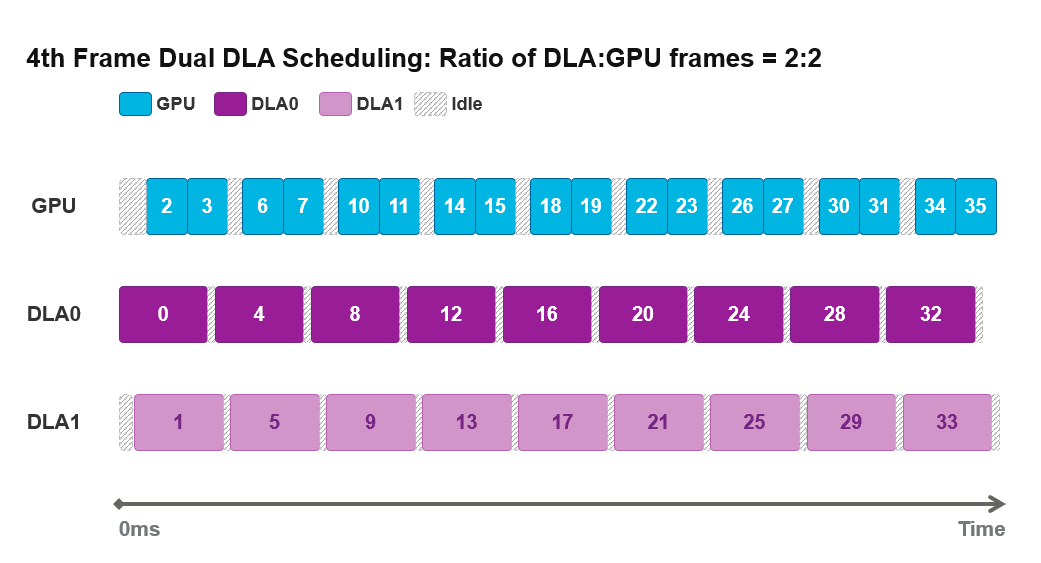}}\hfill
    \subfloat[2:3 ratio. DLA0 receives every 5th frame, and DLA1 receives the subsequent frame; the remaining three frames per cycle go to the GPU. Both DLA cores run concurrently.\label{fig:timing_5th}]{%
        \includegraphics[width=0.48\linewidth]{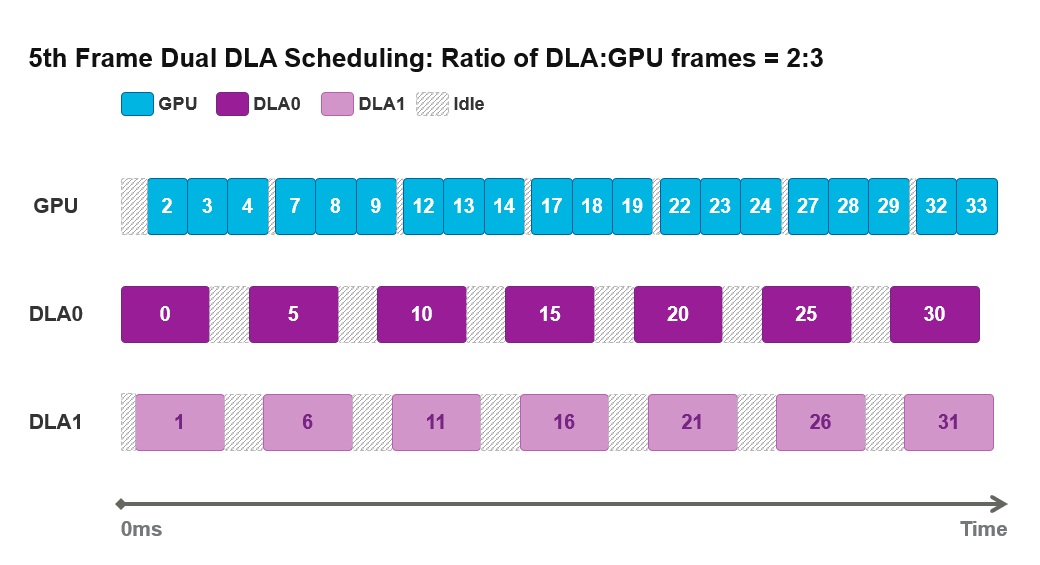}}
    \caption{ \small Timing diagrams for dual-DLA interleaved \hfrads.}
    \label{fig:timing_dual_interleaved}
\end{figure*}

\begin{figure}[H]
    \centering
    \includegraphics[width=1\linewidth]{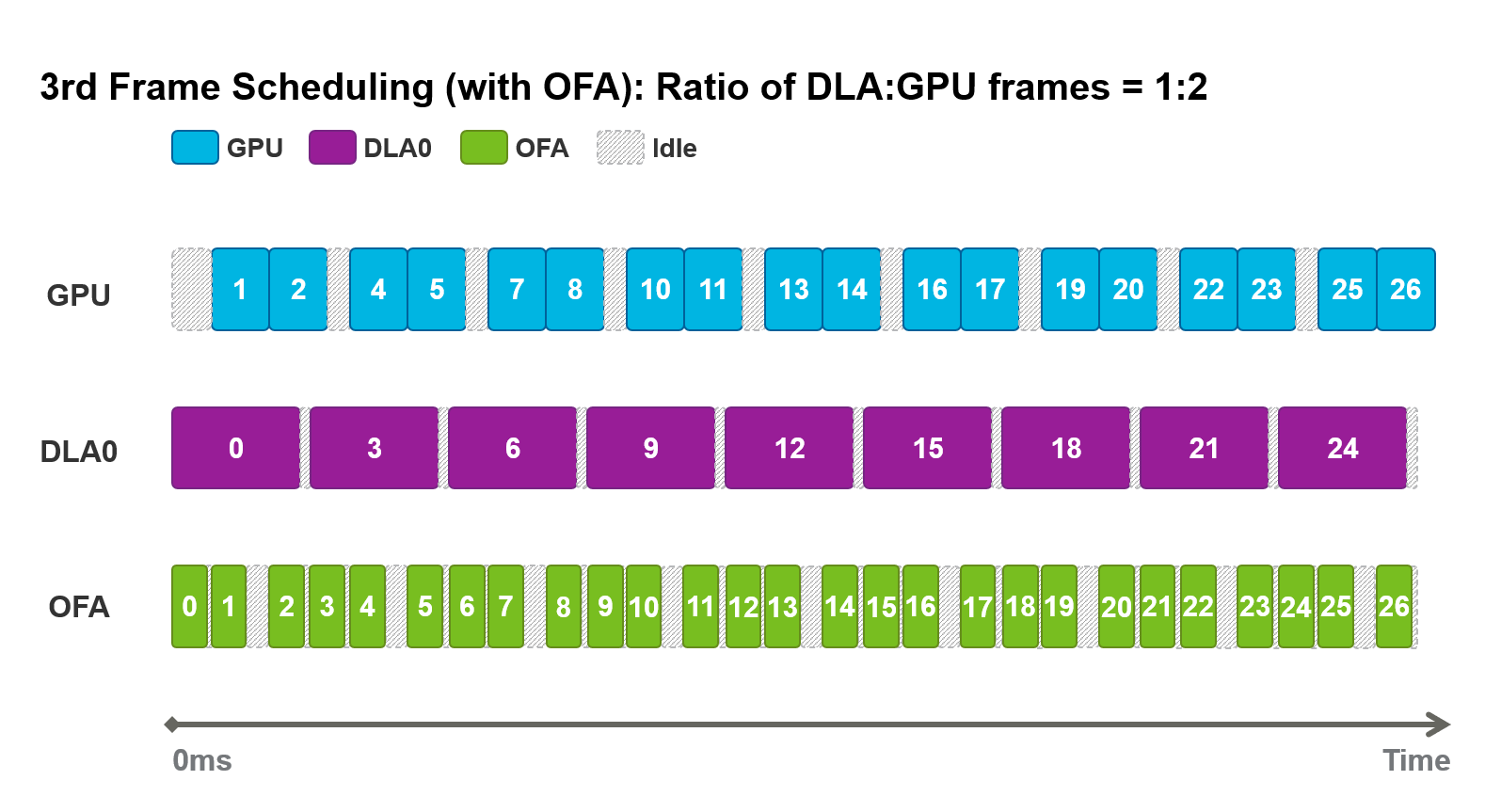}
    \caption{\small Balanced Dispatch (1:2) with \ac{OFA} execution in parallel.}
    \label{fig:3rd_frame_ofa}
\end{figure}

The timing diagrams for the four single-DLA ratios are shown in Fig.~\ref{fig:timing_frame_dispatch}. The 1:1 ratio (Fig.~\ref{fig:even_odd}) illustrates that both units finish well before the next frame arrives, leaving significant idle time on both hardware units. As the ratio increases to 1:2 (Fig.~\ref{fig:timing_3rd}), the DLA latency ($\approx$20 ms) is approximately twice the GPU latency ($\approx$9 ms), achieving near-optimal simultaneous saturation of both units. The same scheduling scheme can also be extended to include \ac{OFA}, enabling concurrent execution across multiple hardware engines, as illustrated in Fig.~\ref{fig:3rd_frame_ofa}. At 1:3 (Fig.~\ref{fig:timing_4th}), the GPU becomes the bottleneck, processing three consecutive frames ($3\times9\approx27$ ms) while the DLA finishes earlier and idles for $\approx$7 ms per cycle. At 1:5 (Fig.~\ref{fig:timing_6th}), the DLA-Sparse configuration leaves the DLA substantially underutilized: after completing its inference at $\approx$20 ms, the next DLA frame does not arrive until 200 ms into the cycle. 

The timing diagrams for the dual-DLA interleaved strategies are shown in Fig.~\ref{fig:timing_dual_interleaved}. In the 2:2 ratio (Fig.~\ref{fig:timing_4th_dual}), DLA0 and DLA1 each receive one frame per four-frame cycle, running concurrently while the GPU handles the remaining two frames. In the 2:3 configuration (Fig.~\ref{fig:timing_5th}), the GPU handles three frames per cycle while both DLA cores together handle two, providing a larger GPU share and a more GPU-dominant balance.

The optimal ratio is model-dependent and is determined by the relative per-inference latencies of the \ac{DLA} and GPU engines. For the proposed model, the \ac{DLA} inference time ($\approx$20 ms) is approximately twice the GPU inference time ($\approx$9 ms), which predicts that the 1:2 Balanced Dispatch ratio will saturate both units simultaneously, a prediction confirmed experimentally in Section~\ref{sec:exp_results}. For the original model, whose \ac{DLA} latency ($\approx$115 ms) is far higher than the GPU's, routing more frames to the GPU monotonically improves total throughput. This illustrates that the optimal dispatch ratio must be re-evaluated whenever architectural modifications alter the per-inference latency balance, thereby directly motivating the evaluation of both model variants across all nine experiments.
 
To the best of the authors' knowledge, \hfrads~is the first frame-level dispatch scheduling strategy to route transformer inference frames across GPU and dual-\ac{DLA} engines on Jetson-class edge GPU platforms.

\begin{figure*}[!t]
    \centering
    \includegraphics[width=0.85\linewidth]{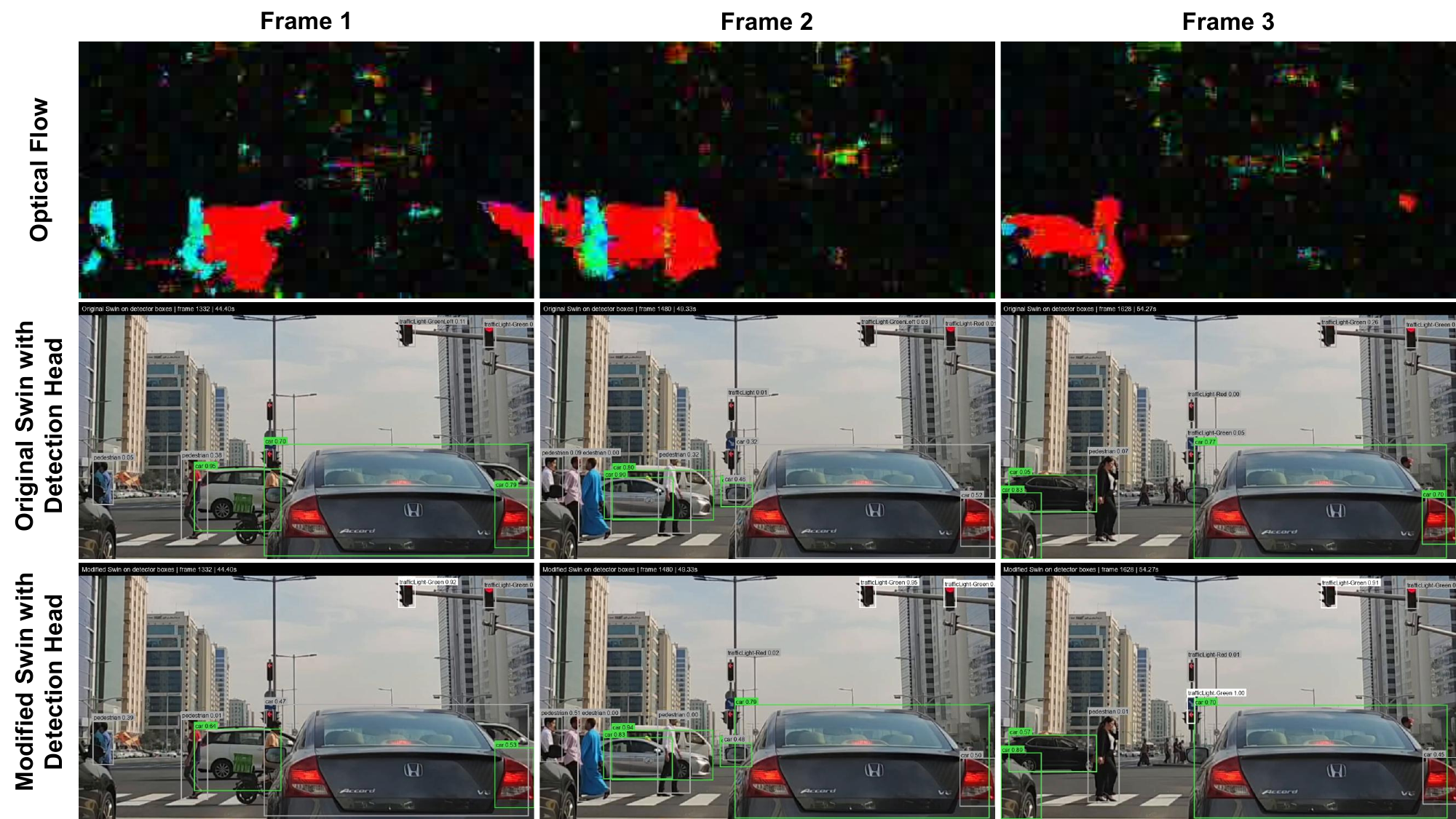}
    \caption{\small Qualitative inference results using \ac{OFA} and the proposed models on daytime road scenes in Abu Dhabi, UAE, across three representative frames.}
    \label{fig:trial_images}
\end{figure*}

\section{Experimental Results and Discussion}
\label{sec:exp_results}

\subsection{Assessment on Dataset}

The primary models were trained and evaluated on the Roboflow Udacity object classification dataset \cite{roboflow}, which provides an autonomous-driving perception benchmark for comparing the original and hardware-aware Swin Transformer variants. For this dataset, the F1 score, the harmonic mean of precision and recall, was used as the primary accuracy metric. The proposed model achieves an F1 score of 92.26\%, compared to 94.91\% for the original model, with an 11.7\% improvement in \ac{DLA}-targeted throughput on the Roboflow Udacity input configuration. The reduction in F1 score is an expected consequence of the architectural substitutions introduced to improve \ac{DLA} compatibility, since these substitutions introduce approximation error during inference. Nevertheless, the proposed model retains most of the original model's discriminative capability, indicating that the hardware-aware modifications preserve the core representational capacity of the Swin Transformer.

To evaluate generalization beyond a single training and validation source, additional experiments were conducted using KITTI and Cityscapes. These datasets were not used with identical task formulations. KITTI was used for evaluation in road/lane bounding-box detection, where the F1 score quantifies detection and localization quality. Cityscapes was used for training and evaluation in a 19-class semantic segmentation task, where pixel accuracy is an appropriate dense-prediction metric. Therefore, the dataset-level results in Table~\ref{tab:dataset_assessment} report the metric corresponding to each dataset's implemented task.

The purpose of these additional experiments is not to claim state-of-the-art accuracy on KITTI or Cityscapes, but to verify that the proposed \ac{DLA}-compatibility substitutions do not overfit to the Roboflow Udacity distribution. Throughput is reported for the measured deployment configuration associated with each dataset: Roboflow Udacity uses the \ac{DLA}-targeted execution configuration, while KITTI and Cityscapes use GPU execution.

\begin{table}[h]
\centering
\caption{\small Dataset-level validation and throughput comparison. Udacity uses DLA-targeted execution; KITTI and Cityscapes use GPU execution.}
\label{tab:dataset_assessment}

\renewcommand{\arraystretch}{0.7}
\setlength{\tabcolsep}{2pt}
\footnotesize

\begin{tabularx}{\columnwidth}{@{}
>{\raggedright\arraybackslash}p{0.20\columnwidth}
>{\centering\arraybackslash}p{0.15\columnwidth}
>{\centering\arraybackslash}p{0.12\columnwidth}
>{\centering\arraybackslash}p{0.12\columnwidth}
>{\centering\arraybackslash}p{0.10\columnwidth}
>{\centering\arraybackslash}p{0.12\columnwidth}
>{\centering\arraybackslash}X
@{}}
\toprule
\multirow{2}{*}{\textbf{Dataset}} &
\multirow{2}{*}{\textbf{Metric}} &
\multicolumn{2}{c}{\textbf{Val. (\%)}} &
\multicolumn{3}{c}{\textbf{FPS}} \\
\cmidrule(lr){3-4}
\cmidrule(lr){5-7}
&
&
\textbf{Orig.} &
\textbf{Prop.} &
\textbf{Dev.} &
\textbf{Orig.} &
\textbf{Prop.} \\
\midrule

Udacity &
F1 &
94.91 & 92.26 &
\ac{DLA} & 47.67 & 53.26 \\

KITTI &
F1@0.5 &
100.00 & 96.55 &
GPU & 260.14 & 262.80 \\

Cityscapes &
Pix. Acc. &
89.11 & 84.84 &
GPU & 209.34 & 199.09 \\

\bottomrule
\end{tabularx}
\vspace{-0.6em}
\end{table}

\vspace{-0.5em}
\subsection{Throughput Performance}
The inference throughput of both models was measured under a fully \ac{DLA}-resident execution configuration on the NVIDIA Jetson AGX Orin, where each model is compiled and run entirely on the \ac{DLA} engine without concurrent GPU inference tasks. Throughput measures the number of frames processed per second, a critical metric for real-time execution pipelines. The corresponding throughput results are included in Table~\ref{tab:dataset_assessment}.
 
The hardware-aware model achieves a throughput of 53.26 FPS, compared to 47.67 FPS for the original model, resulting in an 11\% improvement. This gain is attributed to the reduction in unsupported operations that would otherwise require GPU fallback during inference. The proposed model executes a greater proportion of its computation directly on the \ac{DLA}, reducing the number of GPU-\ac{DLA} transitions and the associated memory transfer overhead. The removal of shuffle layers, which were previously required at format boundaries, further contributes to the observed throughput improvement. A similar difference is also seen in the GPU execution of the models trained on the KITTI and Cityscapes datasets.

\begin{table*}[!t]
\centering
\renewcommand{\arraystretch}{1.2}
\caption{ \small Comparison of concurrent and heterogeneous multi-accelerator scheduling}
\label{tab:naive_ofa_compare}
\begin{tabular}{@{} l l c c c c c c c @{}}
\toprule
\textbf{Scenario} & \textbf{Model} & \textbf{GPU Util (\%)} & \textbf{DLA Util (\%)} & \textbf{OFA Util (\%)} & \textbf{GPU FPS} & \textbf{DLA FPS} & \textbf{OFA FPS} & \textbf{Power (W)} \\
\midrule
 
\multirow{2}{*}{\makecell[l]{Concurrent Task Scheduling}}
& Original & 80.77 & 31.38 & -- & 232.30 & 10.47 & -- & 29.04 \\
& Proposed & 95.68 & 6.18 & -- & 227.89 & 19.67 & -- & 36.67 \\
 
\midrule
 
\multirow{2}{*}{\makecell[l]{Multi-Accelerator Scheduling}}
& Original & 97.30 & 30.50 & 71.05 & 222.46 & 10.26 & 175.75 & 33.87\\
& Proposed & 98.65 & 5.64 & 71.68 & 235.92 & 20.74 & 176.20 & 41.85\\
 
\bottomrule
\end{tabular}
\end{table*}

\subsection{Power and Utilization Measurement Methodology}
\label{sec:power_method}

As energy efficiency is a primary constraint in edge physical AI systems, all power measurements in this work are reported at the system level rather than as isolated accelerator power. Measurements were collected using NVIDIA Tegrastats \cite{tegrastats} during steady-state inference after an initial warm-up. All experiments were performed at a fixed frequency setting: CPU at 2201 MHz, GPU at approximately 1293-1296 MHz, and \ac{DLA} at 1600 MHz, with the \ac{EMC} running at 3199 MHz. Locking the frequency across experiments removes \ac{DVFS} artifacts from the comparison and ensures that observed differences in throughput and power reflect the scheduling architecture rather than operating-point variation. The reported power corresponds to the average platform power during the measurement window, including GPU, \ac{DLA}, memory system, CPU runtime, and scheduler overheads. GPU, \ac{DLA}, and \ac{OFA} utilization are reported as the percentage of time each accelerator remains actively executing inference workloads during the measurement interval.

Throughput per watt is computed as
\begin{equation}
    \eta_{\mathrm{FPS/W}} = \frac{\mathrm{Total\ processed\ frames\ per\ second}}{\mathrm{Average\ platform\ power\ in\ watts}}.
    \label{eq:fps_per_watt}
\end{equation}

For single-model configurations, the numerator corresponds to the Swin Transformer throughput. For concurrent multi-accelerator configurations, the numerator includes the aggregate throughput of simultaneously active perception tasks, because all active hardware engines share the measured power. This distinction is important: multi-task FPS/W evaluates platform utilization efficiency, whereas single-model FPS/W evaluates the efficiency of a specific model deployment.

\subsection{Heterogeneous Scheduling Experiments}
 
To evaluate the throughput and power characteristics of both models under concurrent multi-hardware execution, a set of heterogeneous scheduling strategies was implemented and benchmarked on the NVIDIA Jetson AGX Orin. Timing diagrams for all strategies are presented in Section~\ref{sec:scheduling}.

\subsubsection{\textbf{Concurrent Task Scheduling (YOLOv8 on GPU and Swin Transformer on DLA) and Multi-Accelerator Scheduling (GPU, DLA, and \ac{OFA})}}
 
In this configuration, a YOLOv8 object detection model runs concurrently on the GPU. In contrast, the Swin Transformer runs on \ac{DLA} core 0, each in an independent thread with its own event-synchronized inference loop. Table~\ref{tab:naive_ofa_compare} reports the results under the fixed frequency setting.
 
The proposed model achieves 19.67 DLA FPS, compared to 10.47 FPS for the original, representing an 88\% increase in DLA throughput. GPU throughput remains comparable between the two models (232.30 vs. 227.89 FPS), as the YOLOv8 workload is identical in both cases. The lower \ac{DLA} utilization of the proposed model (6.18\% vs. 31.38\%) reflects faster per-inference execution: the \ac{DLA} completes each frame more quickly and spends proportionally less time active. The increase in total power (29.04 vs. 36.67 W) is primarily driven by higher GPU utilization (95.68\% vs. 80.77\%), resulting from the faster \ac{DLA}, which releases shared memory resources sooner and allows the GPU to sustain a higher throughput.
 
In the heterogeneous three-task configuration, YOLOv8 runs on the GPU, the Swin Transformer runs on \ac{DLA} core 0, and a dense optical flow model runs on the \ac{OFA} hardware accelerator, all concurrently in independent threads. Table~\ref{tab:naive_ofa_compare} reports the results. The timing diagram is shown in Fig.~\ref{fig:timing_ofa}.
 
The proposed model achieves 20.74 DLA FPS, compared with 10.26 FPS for the original, consistent with the standalone DLA speedup. GPU and \ac{OFA} throughput are comparable between the two models, as neither the YOLOv8 model nor the optical flow engine changes. Adding the \ac{OFA} task increases EMC bus contention relative to the naive two-task case, slightly suppressing \ac{DLA} throughput for both models compared to the standalone Naive Scheduling results.

\subsubsection{\textbf{\hfrads: Alternate, Balanced, GPU-Dominant, and DLA-Sparse}}

\hfrads~routes every N-th frame to the \ac{DLA} and the remaining $N-1$ frames to the GPU per cycle, using the same non-blocking concurrent design. Four ratios were evaluated: 2nd frame (1:1 DLA:GPU), 3rd frame (1:2 DLA:GPU), 4th frame (1:3 DLA:GPU), and 6th frame (1:5 DLA:GPU). Timing diagrams are shown in Fig.~\ref{fig:timing_frame_dispatch} in Section~\ref{sec:scheduling}. Table~\ref{tab:nthframe_freq} reports the utilization and power results for the 3rd, 4th, and 6th frame strategies. To further isolate the parallel \ac{OFA}-enabled 3rd-frame configuration, Table~\ref{tab:hfrads_ofa_third_frame} summarizes the CPU, GPU, \ac{DLA}, throughput, and power behavior when \hfrads~runs with \ac{OFA} execution in parallel.

\begin{table}[H]
\centering
\renewcommand{\arraystretch}{1.15}
\setlength{\tabcolsep}{3pt}
\caption{ \small Utilization and power consumption of \hfrads~strategies}
\label{tab:nthframe_freq}
\begin{tabular}{@{} l l c c c @{}}
\toprule
\textbf{Strategy} & \textbf{Model} &
\textbf{GPU} & \textbf{DLA} & \textbf{Power} \\
& & \textbf{Util. (\%)} & \textbf{Util. (\%)} & \textbf{(W)} \\
\midrule
\multirow{2}{*}{2nd Frame}
  & Original  & 32.48 & 48.08 & 21.08 \\
  & Proposed  & 94.40 & 16.54 & 31.30 \\
\midrule
\multirow{2}{*}{3rd Frame}
  & Original  & 37.23 & 49.14 & 23.23 \\
  & Proposed  & 95.93 &  7.47 & 31.15 \\
\midrule
\multirow{2}{*}{4th Frame}
  & Original  & 45.94 & 46.36 & 21.96 \\
  & Proposed  & 94.25 &  4.56 & 29.40 \\
\midrule
\multirow{2}{*}{6th Frame}
  & Original  & 63.24 & 48.70 & 24.80 \\
  & Proposed  & 94.38 &  0.30 & 32.16 \\
\bottomrule
\end{tabular}
\end{table}

\begin{table}[H]
\centering
\setlength{\tabcolsep}{3pt}
\renewcommand{\arraystretch}{1.15}
\caption{\small 3rd-frame OFA scheduling resource summary}
\label{tab:hfrads_ofa_third_frame}
\begin{tabular}{@{}lcccccc@{}}
\toprule
\textbf{Model} &
\makecell{\textbf{GPU}\\\textbf{Util. (\%)}} &
\makecell{\textbf{DLA}\\\textbf{Util. (\%)}} &
\makecell{\textbf{DLA}\\\textbf{FPS}} &
\makecell{\textbf{GPU}\\\textbf{FPS}} &
\makecell{\textbf{OFA}\\\textbf{FPS}} &
\makecell{\textbf{Power}\\\textbf{(W)}} \\
\midrule
Original & 34.68 & 45.41 & 17.42 & 34.84 & 52.26 & 24.45 \\
Proposed & 93.09 & 8.32 & 41.75 & 83.51 & 125.26 & 36.07 \\
\bottomrule
\end{tabular}
\end{table}

\begin{figure}[ht]
    \centering
    \includegraphics[width=\linewidth]{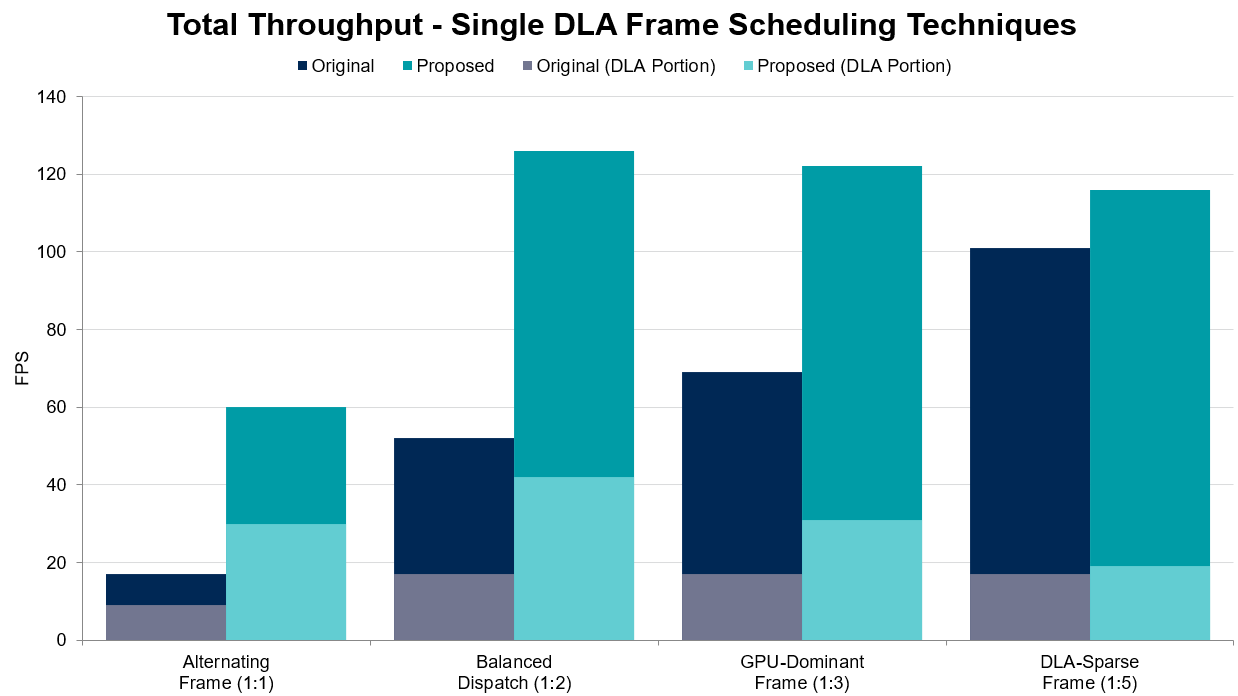}
    \caption{ \small Throughput comparison between the different \hfrads~strategies.}
    \label{fig:throughput_graph_frame}
\end{figure}

Across all \hfrads~strategies, total FPS increases with the GPU:DLA ratio for the original model, since the original \ac{DLA} is the bottleneck (per-inference latency $\approx$115 ms) and routing more frames to the faster GPU raises throughput. For the proposed model, total FPS peaks at the 3rd frame ratio (125.93 FPS) in free-running mode and declines at higher ratios. The proposed model's \ac{DLA} latency ($\approx$20 ms) is close to the GPU latency ($\approx$9 ms), indicating an optimal 1:2 DLA:GPU ratio for equal hardware loading, consistent with the 3rd-frame strategy delivering the highest free-running throughput. At a paced input rate, however, the input frame rate equally limits all \hfrads~strategies, and the choice of ratio affects only the hardware idle distribution rather than total system throughput. In the 2nd-frame \hfrads~strategy, the proposed model achieves 60.37 total FPS vs. 17.35 FPS for the original, a 248\% improvement. The low \ac{DLA} utilization of the proposed model (16.54\%) is consistent with its faster per-frame \ac{DLA} inference, which completes well within the time budget available in each 1:1 cycle. The original model has a higher \ac{DLA} utilization (48.08\%), reflecting the longer per-inference time of the unmodified engine. The total throughput comparison across all single-DLA \hfrads~strategies is shown in Fig.~\ref{fig:throughput_graph_frame}. With parallel \ac{OFA} execution, the 3rd-frame \hfrads~strategy preserves the proposed model's throughput advantage and produces results similar to those of the proposed model, achieving 41.75 \ac{DLA} FPS and 83.51 GPU FPS while maintaining moderate power consumption at 36.07 W.

\subsubsection{\textbf{Symmetric DLA Scheduling: Two Swin Transformer Instances on DLA0 and DLA1}}
 
In this configuration, two independent instances of the Swin Transformer run concurrently, one on each \ac{DLA} core, without any GPU inference task. Table~\ref{tab:dla_freq} reports the results. The timing diagram is shown in Fig.~\ref{fig:timing_dual_dla} in Section~\ref{sec:scheduling}.

\begin{table}[H]
\centering
\setlength{\tabcolsep}{3pt}
\renewcommand{\arraystretch}{1.15}
\caption{ \small Symmetric DLA scheduling (Swin on DLA0 and DLA1)}
\label{tab:dla_freq}
\begin{tabular}{@{}lcccccc@{}}
\toprule
\textbf{Model} &
\makecell{\textbf{GPU}\\\textbf{Util. (\%)}} &
\makecell{\textbf{DLA0}\\\textbf{Util. (\%)}} &
\makecell{\textbf{DLA1}\\\textbf{Util. (\%)}} &
\makecell{\textbf{DLA0}\\\textbf{FPS}} &
\makecell{\textbf{DLA1}\\\textbf{FPS}} &
\makecell{\textbf{Power}\\\textbf{(W)}} \\
\midrule
Original & 57.97 & 44.07 & 47.92 & 15.97 & 15.95 & 29.15 \\
Proposed & 88.19 & 15.91 & 16.17 & 57.22 & 57.15 & 37.60 \\
\bottomrule
\end{tabular}
\end{table}
 
Both \ac{DLA} cores show near-identical throughput across models, confirming that the two engines run symmetrically. The proposed model achieves 57.22 and 57.15 FPS on DLA0 and DLA1, respectively, compared to 15.97 and 15.95 FPS for the original, a 3.6$\times$ improvement per core. Since no explicit GPU task is scheduled, the observed GPU utilization (57.97\% original, 88.19\% proposed) is entirely due to GPU fallback operations executed within the TensorRT \ac{DLA} engines. The higher GPU utilization for the proposed model reflects the fact that although the total number of GPU-fallback layers is reduced, the remaining fallback operations are individually more GPU-intensive, as they represent the operations most incompatible with the \ac{DLA}.

\begin{table}[ht]
\centering
\setlength{\tabcolsep}{3pt}
\renewcommand{\arraystretch}{1.15}
\caption{ \small Comparison of dual-DLA interleaved \hfrads~strategies}
\label{tab:dispatch_dual_dla}
\begin{tabular}{@{}llcccc@{}}
\toprule
\textbf{Scenario} &
\textbf{Model} &
\makecell{\textbf{GPU}\\\textbf{Util. (\%)}} &
\makecell{\textbf{DLA0}\\\textbf{Util. (\%)}} &
\makecell{\textbf{DLA1}\\\textbf{Util. (\%)}} &
\makecell{\textbf{Power}\\\textbf{(W)}} \\
\midrule
\multirow{2}{*}{4th Frame (2:2)}
& Original & 66.52 & 40.34 & 47.50 & 32.45 \\
& Proposed & 82.85 & 1.52 & 1.81 & 36.24 \\
\midrule
\multirow{2}{*}{5th Frame (2:3)}
& Original & 69.11 & 48.26 & 47.51 & 23.65 \\
& Proposed & 85.48 & 0.33 & 0.98 & 27.32 \\
\bottomrule
\end{tabular}
\end{table}

\begin{figure}[h]
    \centering
    \includegraphics[width=1\linewidth]{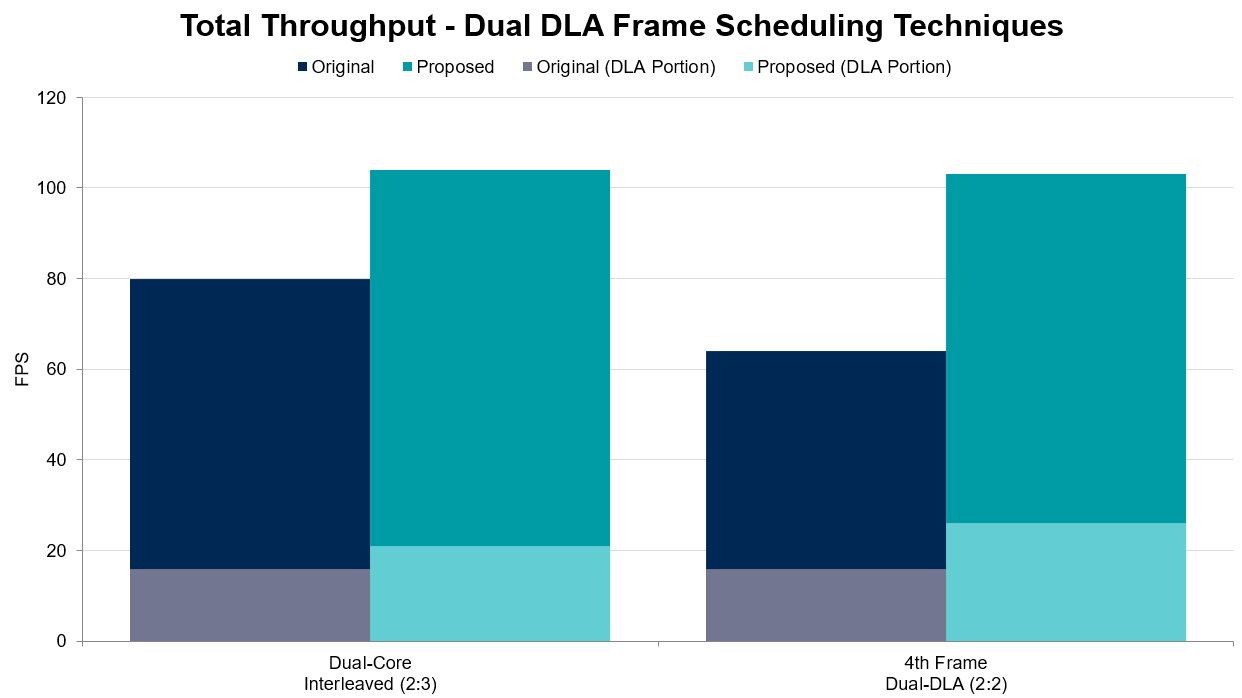}
    \caption{ \small Throughput comparison between the different dual DLA scheduling strategies.}
    \label{fig:throughput_dual_graph}
\end{figure}

\subsubsection{\textbf{Dual-DLA Interleaved \hfrads}}
 
This methodology extends single-\ac{DLA} \hfrads. It routes two consecutive frames to DLA0 and DLA1, respectively, per Nth-frame cycle ($\text{frameId} \bmod N = 0 \rightarrow \text{DLA0}$; $\text{frameId} \bmod N = 1 \rightarrow \text{DLA1}$; remainder $\rightarrow \text{GPU}$). Two ratios were tested: 4th (2:2 DLA:GPU) and 5th (2:3 DLA:GPU). Table~\ref{tab:dispatch_dual_dla} reports the utilization and power results. The timing diagrams are shown in Fig.~\ref{fig:timing_dual_interleaved} in Section~\ref{sec:scheduling}. Both ratios offer similar overall throughput, with the 5th-frame strategy yielding 104.04 FPS and the 4th-frame strategy yielding 103.30 FPS (using the proposed transformer model variant). The throughput comparison across the evaluated configurations is summarized in Fig.~\ref{fig:throughput_dual_graph}.

\vspace{-1em}
\subsection{Latency and Power Efficiency Analysis}
 
A critical requirement for any autonomous driving perception system is that inference latency remains within the bounds imposed by the vehicle's reaction budget. At 30 FPS camera input, the maximum tolerable frame-to-frame latency is 33.3 ms (one frame period). Although ISO 26262 does not mandate fixed perception latency thresholds, automotive perception pipelines are commonly designed around real-time constraints near 30 FPS ($\approx$ 33 ms). In this work, we additionally evaluate latency regimes relevant to urban driving ($\leq$ 50 ms) and highway driving ($\leq$ 100 ms), corresponding to object displacements of approximately 0.7 m at 50 km/h and 3.6 m at 130 km/h per inference cycle \cite{iso}.

\begin{figure}[ht]
    \centering
    \includegraphics[width=0.9\linewidth]{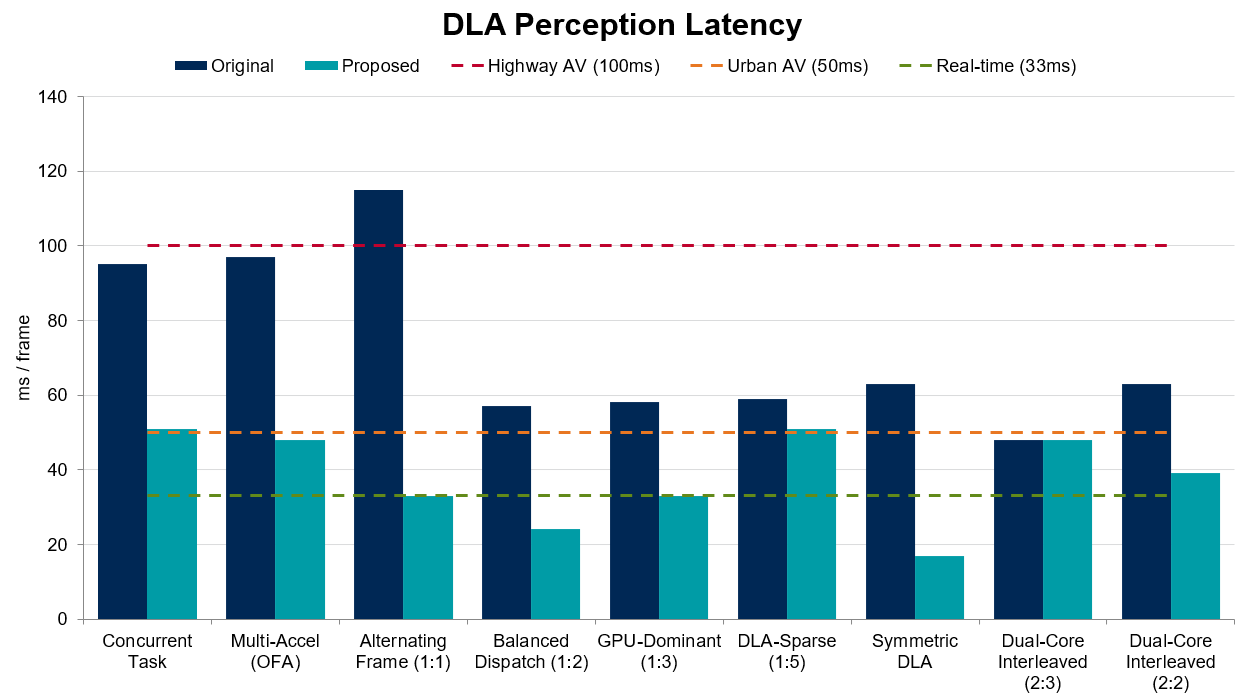}
    \caption{ \small \ac{DLA} perception latency (ms/frame) for each scheduling strategy, original and proposed models. Dashed lines mark the real-time 30 FPS threshold (33 ms), the urban \ac{AV} budget (50 ms), and the highway \ac{AV} budget (100 ms).}
    \label{fig:latency_av}
\end{figure}

Fig.~\ref{fig:latency_av} reports the per-frame \ac{DLA} perception latency (reciprocal of \ac{DLA} FPS) for every scheduling strategy evaluated in this work. The three automotive thresholds are overlaid as dashed reference lines. For the original model, only the Dual-Core Interleaved (2:3) strategy falls below the 50-ms urban threshold. In comparison, most strategies exceed the 100 ms highway budget, particularly the Concurrent Task and Multi-Accelerator strategies, where \ac{DLA} latency reaches $\approx$96 ms and $\approx$97 ms, respectively, due to the high per-inference cost of the unmodified engine. The proposed model substantially changes this picture: every single-\ac{DLA} \hfrads~strategy (Alternating Frame through DLA-Sparse) and both Symmetric DLA configurations achieve latencies below 50 ms, with Balanced Dispatch (1:2) achieving $\approx$24 ms, comfortably within the real-time 30 FPS budget. The Dual-Core Interleaved and Symmetric DLA strategies achieve $\approx$48 ms and $\approx$9 ms, respectively, for the proposed model. Even the Concurrent Task strategy, which operates in free-running mode, achieves $\approx$51 ms with the proposed model, narrowly missing the urban threshold. These results confirm that the proposed \ac{DLA}-compatible Swin Transformer is suitable for real-time autonomous driving perception across a range of heterogeneous scheduling configurations. In contrast, the original model fails to meet urban or real-time latency budgets across all tested strategies.

\begin{figure}[ht]
    \centering
    \includegraphics[width=0.9\linewidth]{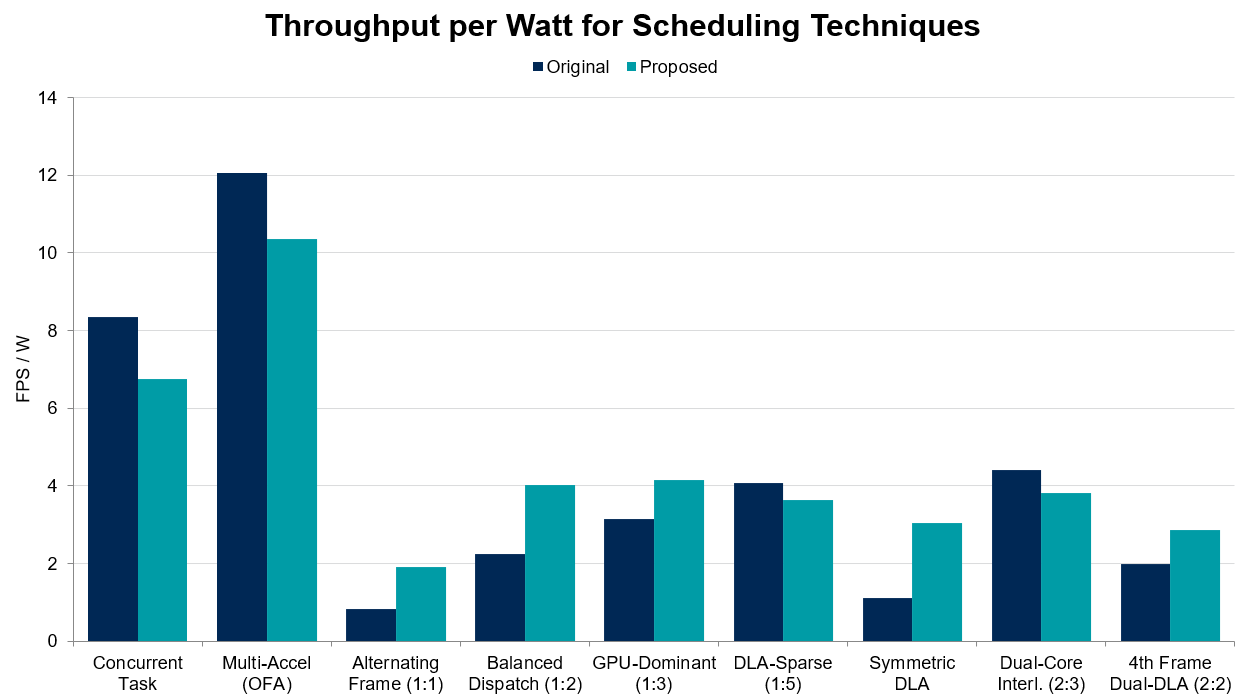}
    \caption{ \small System throughput per watt (FPS/W) across all scheduling strategies. Higher values indicate better power efficiency.}
    \label{fig:throughput_per_watt}
\end{figure}

Fig.~\ref{fig:throughput_per_watt} reports the system-level throughput per watt (FPS/W) for all strategies. The Heterogeneous Multi-Accelerator strategy achieves the highest FPS/W for both models (original: 12.1, proposed: 10.3 FPS/W), as the three concurrent tasks amortize the system power draw across the largest total frame count. Among single-\ac{DLA} \hfrads~strategies, Balanced Dispatch yields the best FPS/W for the proposed model (4.0 FPS/W), reflecting its near-optimal hardware loading. The Symmetric DLA strategy achieves only 3.0 FPS/W, despite high per-core throughput, because it draws full system power when no GPU inference task is running. The highest aggregate FPS/W is observed in the multi-accelerator configuration because GPU, \ac{DLA}, and \ac{OFA} workloads share the same platform power budget. However, this should not be interpreted as the most efficient single-model Swin deployment. Among the \hfrads~single-model dispatch configurations, Balanced Dispatch provides the best latency-efficiency trade-off for the proposed Swin model, achieving $\approx$24 ms latency and 4.0 FPS/W.
 
\begin{table*}[!t]
\centering
\caption{\small Positioning of the proposed method relative to representative edge-transformer optimization approaches.}
\label{tab:sota_positioning}

\renewcommand{\arraystretch}{0.8}
\setlength{\tabcolsep}{2.2pt}

\begin{tabularx}{\textwidth}{@{}
>{\raggedright\arraybackslash}p{0.18\textwidth}
>{\raggedright\arraybackslash}p{0.15\textwidth}
>{\raggedright\arraybackslash}p{0.17\textwidth}
>{\raggedright\arraybackslash}p{0.22\textwidth}
>{\raggedright\arraybackslash}X
@{}}
\toprule

\textbf{Method} &
\textbf{Target Platform} &
\textbf{Optimization Type} &
\textbf{Transformer Compatibility Handling} &
\textbf{Main Difference from This Work} \\

\midrule

Map-and-Conquer~\cite{bouzidi2023mapandconquer} &
Jetson AGX Xavier GPU/\ac{DLA} &
Network partitioning and heterogeneous mapping &
Assumes target layers are accelerator-compatible &
Improves scheduling after compatibility is available, but does not modify incompatible transformer operators. \\

FPGA Swin accelerator~\cite{liu2023efficientfpgabasedacceleratorswin} &
Custom FPGA &
Hardware--software co-design &
Replaces LayerNorm and approximates nonlinear functions &
Requires a custom accelerator design, whereas this work targets commercial NVIDIA Jetson \ac{DLA}/GPU/\ac{OFA} hardware. \\

Compression-based edge ViT methods~\cite{vit_edge_survey2025} &
General edge devices &
Quantization, pruning, and distillation &
Usually preserves the original operator graph &
Reduces model size or arithmetic cost, but may not remove \ac{DLA}-unsupported operators. \\

\midrule

\textbf{Proposed method} &
\textbf{NVIDIA Jetson AGX Orin GPU/\ac{DLA}/\ac{OFA}} &
\textbf{Operator adaptation and frame-level heterogeneous scheduling} &
\textbf{Reshapes 3D tensors to 4D, replaces \ac{ERF}-based \ac{GELU}, and substitutes LayerNorm with bounded $\tanh$} &
\textbf{Targets practical edge physical AI deployment on existing heterogeneous \ac{AI-SoC} hardware.} \\

\bottomrule
\end{tabularx}

\vspace{-0.8em}
\end{table*}
 
When analyzing the trade-offs, the proposed model strategies lie in the low-latency, moderate-efficiency region ($<$50 ms, 2-4 FPS/W). The \hfrads~Balanced Dispatch strategy occupies the most favorable position among the \hfrads~group: 24 ms latency and 4.0 FPS/W efficiency. No original-model strategy achieves both latency $<$100 ms and efficiency $>$2 FPS/W simultaneously. The proposed model therefore dominates across the latency-efficiency Pareto frontier, making it the preferred configuration for real-time autonomous driving deployments on the NVIDIA Jetson AGX Orin.
 
\vspace{-1em} 
\section{Discussion and Comparison with Other Related Works}

The results show that transformer-based vision models can be adapted for heterogeneous edge-AI platforms with dedicated \acp{DLA}, GPU resources, and \ac{OFA}. By replacing unsupported operators, restructuring incompatible tensor formats, and exploiting concurrent execution, the proposed methodology improves throughput on commercially available edge hardware. The \hfrads~Balanced Dispatch configuration further improves utilization by distributing frames according to accelerator latency, reducing idle periods, and increasing system throughput.

The proposed method differs from prior optimization approaches because it intervenes directly in the deployment stack. Compression methods such as quantization, pruning, and distillation reduce arithmetic cost, but do not necessarily make the graph executable on fixed-function accelerators. Map-and-Conquer improves heterogeneous mapping, but assumes that the selected hardware engine already supports the mapped operators. FPGA-based Swin accelerators jointly modify the algorithm and the hardware but require a custom accelerator fabric. In contrast, this work targets commercial Jetson-class \acp{AI-SoC} and combines graph-level compatibility adaptation with frame-level scheduling across GPU, \ac{DLA}, and \ac{OFA}.

Compared with prior work, this contribution complements edge-AI optimization and transformer-acceleration studies. Quantization and algorithm-level methods improve efficiency on resource-constrained platforms~\cite{li2022quantizationedge}, while specialized transformer accelerators accelerate softmax, nonlinear functions, or FPGA-based transformer execution~\cite{fu2024softact,10623817,11556200}. However, these approaches typically require algorithmic redesign or custom hardware. The proposed methodology instead enables deployment on existing heterogeneous edge-GPU platforms by adapting the neural network architecture to satisfy \ac{DLA} constraints while maximizing concurrency across available engines.

This deployment focus is important for autonomous driving and edge-AI systems, where multiple perception tasks must run simultaneously under latency, power, and thermal constraints~\cite{deng2021deeplearningADS}. The \ac{DLA} is attractive because it provides efficient execution for structured inference workloads while freeing the GPU for more complex computations. However, current \ac{DLA} architectures remain optimized mainly for convolutional workloads, whereas vision transformers include operators such as \ac{GELU}, LayerNorm, and large matrix multiplications that trigger GPU fallback.

Previous heterogeneous-scheduling studies primarily address workload orchestration and resource allocation across accelerators~\cite{bouzidi2023mapandconquer,mri_pipeline,hu2024microservices,review}. They generally assume that the target model is already executable on the selected engine. This work addresses the preceding compatibility problem by transforming the model before scheduling, thereby reducing fallback-induced fragmentation and enabling more effective use of GPU, \ac{DLA}, and \ac{OFA} resources. 

Overall, the presented approach bridges the gap between modern transformer architectures and current heterogeneous edge-AI hardware. Rather than relying on custom transformer accelerators, which are not yet widely available in commercial edge-GPU platforms, it adapts transformer models to existing accelerator constraints and schedules them across available engines. This provides a practical pathway for deploying transformer-based perception on embedded \acp{AI-SoC}. The proposed operator substitutions are hardware-driven approximations and are not mathematically equivalent to the original Swin Transformer operations. In particular, the bounded $\tanh$ surrogate for LayerNorm preserves activation range control but does not reproduce exact per-sample centering or variance normalization. Therefore, the method requires task-level validation after substitution and may require calibration for other transformer architectures or datasets. In addition, power is measured at the platform level, which reflects practical deployment cost but does not isolate per-engine energy consumption. Finally, \hfrads~uses deterministic fixed dispatch ratios; adaptive scheduling based on queue length, thermal state, or power budget could further improve robustness under dynamic workloads.

\begin{figure}[H]
        \centering
        \includegraphics[width=0.9\linewidth]{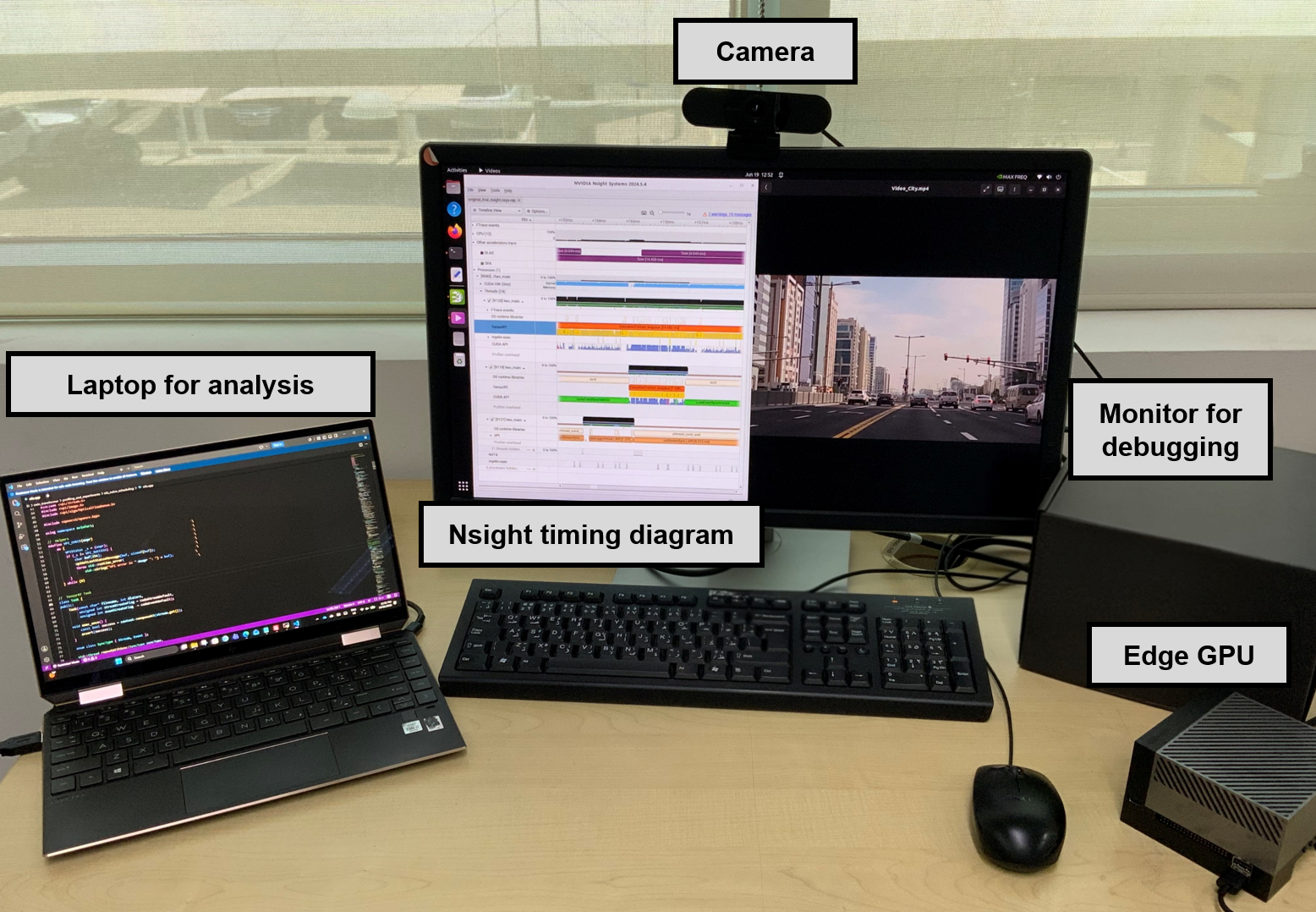}
        \caption{\small Experimental Setup.}
        \label{fig:experimental_setup}
\end{figure}

\vspace{-2em}
\section{Conclusion}
This paper presents an edge physical AI deployment methodology for adapting transformer-based vision models to execute efficiently on heterogeneous NVIDIA Jetson-class \ac{AI-SoC} platforms, demonstrating it with the Swin Transformer on the Jetson AGX Orin. Three architectural substitutions (4D tensor reshaping, \ac{ERF}-to-$\tanh$ substitution in \ac{GELU}, and layer normalization replacement with a $\tanh$-based non-linearity) eliminate the \ac{DLA}-incompatible operations that are structurally common across hierarchical vision transformer architectures, yielding an 11\% throughput improvement (47.67 to 53.26 FPS) at a 2.65\% F1 score cost on the Roboflow Udacity dataset. Five heterogeneous scheduling strategies were benchmarked across the Jetson's GPU, dual \ac{DLA} cores, and, for the first time, the \ac{OFA} as an active concurrent inference unit. The \hfrads~Balanced Dispatch (1:2) experiment achieves 125.93 FPS, a 2.5$\times$ gain over standalone \ac{DLA} operation, by matching the dispatch ratio to the hardware latency balance. All nine experiments meet the 50~ms urban driving budget with the proposed model, and three satisfy the 33~ms real-time threshold, with \hfrads~Balanced Dispatch identified as the Pareto-optimal point at 24~ms and 4.0~FPS/W. These results establish that hardware-aware architectural adaptation of vision transformers, combined with optimized heterogeneous scheduling across all available on-chip accelerators, is a practical and effective pathway to deploying state-of-the-art perception models on embedded edge \ac{AI-SoC} platforms.

\vspace{-1.25em}

\ifCLASSOPTIONcaptionsoff
  \newpage
\fi

\begingroup
\bibliographystyle{IEEEtran}
\bibliography{bibtex/bib/IEEEref}
\endgroup

\end{document}